\documentclass[sigconf,natbib=false]{acmart}

\usepackage[table]{xcolor}
\usepackage{cleveref}
\usepackage{multirow}
\usepackage{tablefootnote}
\usepackage{float}
\usepackage[flushleft]{threeparttable}
\usepackage{graphicx}
\usepackage{bbm}
\usepackage{bm}
\usepackage[justification=centering]{caption}

\usepackage{subcaption}

\usepackage{tikz}
\usetikzlibrary{positioning, arrows.meta, shapes, calc, matrix, shadows}
\usetikzlibrary{decorations.pathreplacing}
\usetikzlibrary{external}
\usetikzlibrary{tikzmark}
\usetikzlibrary{external}
\AtBeginDocument{%
}

\copyrightyear{2025}
\acmYear{2025}
\setcopyright{acmlicensed}\acmConference[SIGIR '25]{Proceedings of the 48th
International ACM SIGIR Conference on Research and Development in
Information Retrieval}{July 13--18, 2025}{Padua, Italy}
\acmBooktitle{Proceedings of the 48th International ACM SIGIR Conference on
Research and Development in Information Retrieval (SIGIR '25), July 13--18,
2025, Padua, Italy}
\acmDOI{10.1145/3726302.3729904}
\acmISBN{979-8-4007-1592-1/2025/07}

\RequirePackage[
  datamodel=acmdatamodel,
  style=acmnumeric,
  ]{biblatex}

\begin{document}

\newcommand{\WARP}{WARP}

\title{WARP: An Efficient Engine for Multi-Vector Retrieval}

\author{Jan Luca Scheerer}
\authornote{Work completed as a visiting student researcher at Stanford University.}
\affiliation{%
  \institution{ETH Zurich}
  \country{Zurich, Switzerland}
}
\email{lscheerer@ethz.ch}

\author{Matei Zaharia}
\affiliation{%
  \institution{UC Berkeley}
  \country{Berkeley, CA, USA}}
\email{matei@berkeley.edu}

\author{Christopher Potts}
\affiliation{%
  \institution{Stanford University}
  \country{Stanford, CA, USA}
}
\email{cgpotts@stanford.edu}

\author{Gustavo Alonso}
\affiliation{%
 \institution{ETH Zurich}
 \country{Zurich, Switzerland}}
 \email{alonso@inf.ethz.ch}

\author{Omar Khattab}
\affiliation{%
  \institution{Stanford University}
  \country{Stanford, CA, USA}}
\email{okhattab@cs.stanford.edu}

\renewcommand{\shortauthors}{Scheerer et al.}

\begin{abstract}
Multi-vector retrieval methods such as ColBERT and its recent variant, the ConteXtualized Token Retriever (XTR), offer high accuracy but face efficiency challenges at scale. To address this, we present WARP, a retrieval engine that substantially improves the efficiency of retrievers trained with the XTR objective through three key innovations: (1) \WARP{}$_\text{SELECT}$ for dynamic similarity imputation; (2) implicit decompression, avoiding costly vector reconstruction during retrieval; and (3) a two-stage reduction process for efficient score aggregation. Combined with highly-optimized C++ kernels, our system reduces end-to-end latency compared to XTR's reference implementation by 41x, and achieves a 3x speedup over the ColBERTv2/PLAID engine, while preserving retrieval quality.

\vspace{0.6em}
\hspace{1.8em}\includegraphics[width=1.25em,height=1.25em]{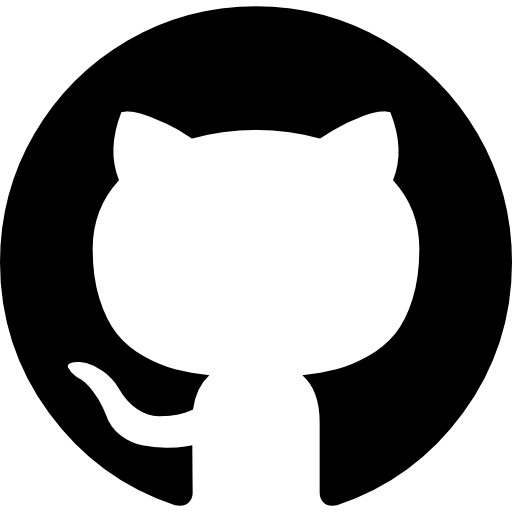}\hspace{.3em}
\parbox[c]{\columnwidth}
{
    \vspace{-.55em}
    \href{https://github.com/jlscheerer/xtr-warp}{\nolinkurl{https://github.com/jlscheerer/xtr-warp}}
}
\vspace{-1.2em}

\end{abstract}


\maketitle

\section{Introduction}

Over the past several years, information retrieval (IR) research has introduced new neural paradigms for search based on pretrained Transformers. Central among these, the late interaction paradigm proposed in ColBERT \cite{colbert} departs from the bottlenecks of conventional \textit{single-vector} representations. Instead, it encodes queries and documents into \textit{multi-vector} representations on top of which it is able to scale gracefully to search massive collections.

Since the original ColBERT architecture was introduced, there has been substantial research in optimizing the latency of multi-vector retrieval models \cite{coil, citadel, emvb}. Most notably, PLAID \cite{plaid} reduces late interaction search latency by $45x$ on a CPU compared to a vanilla ColBERTv2~\cite{colbert2} process, while continuing to deliver state-of-the-art retrieval quality. Orthogonally, the ConteXtualized Token Retriever (XTR) \cite{xtr} introduces a novel training objective that eliminates the need for a separate gathering stage and thereby significantly simplifies the subsequent scoring stage. While XTR\footnote{\url{https://github.com/google-deepmind/xtr}} lays extremely promising groundwork for more efficient multi-vector retrieval, we find that it naively relies on a general-purpose vector similarity search library (ScaNN) and combines that with unoptimized Python data structures and manual iteration, introducing substantial overhead.

\begin{figure}[t]
    \centering
    \includegraphics[width=0.45\textwidth]{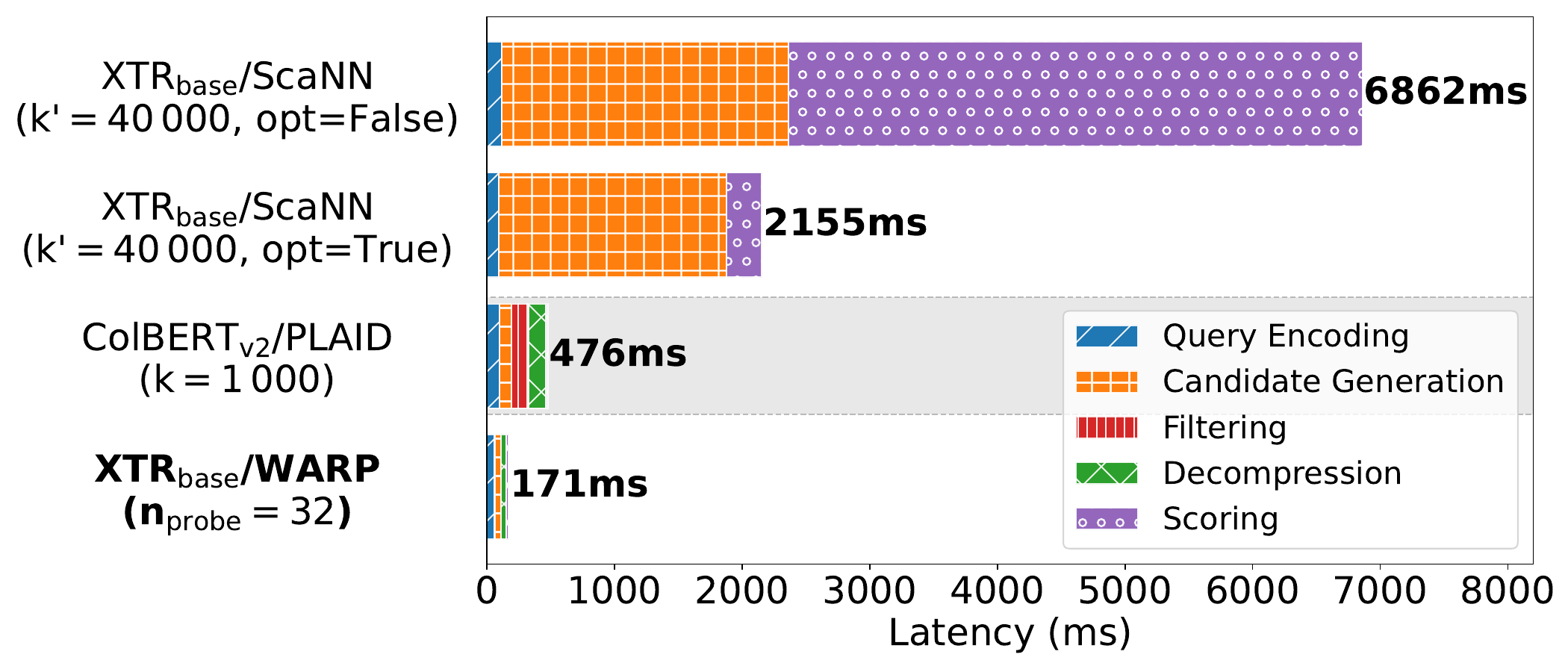}
    \caption[Latency breakdown comparison on LoTTE Pooled]{Single-threaded CPU latency breakdown of (1) XTR's unoptimized reference implementation, (2) a variant of XTR that we optimized, (3) the official ColBERTv2/PLAID system, and (4) our proposed XTR/\WARP{} on LoTTE Pooled.}
    \label{fig:comp_lotte_pooled_test}
\end{figure}

The key insights from PLAID and XTR appear rather isolated. Whereas PLAID is concerned with aggressively and swiftly pruning away documents it finds unpromising, XTR tries to eliminate gathering complete document representations in the first place. We ask whether there are potentially rich interactions between these two fundamentally distinct approaches to speeding up multi-vector search. To study this, we introduce a new engine for retrieval with XTR-based ColBERT models, called \WARP{}, that combines techniques from ColBERTv2/PLAID with innovations tailored for the XTR architecture. Our contributions in \WARP{} include: (1) the WARP$_\text{SELECT}$ method for imputing missing similarities, (2) a new method for implicit decompression of vectors during search, and (3) a novel two-stage reduction phase for efficient scoring.

Experimental evaluation shows that \WARP{} achieves a 41x reduction in end-to-end latency compared to the XTR reference implementation on LoTTE Pooled, bringing query response times down from above 6 seconds to just 171 milliseconds in single-threaded execution, while also reducing index size by a factor of 2x--4x compared to the ScaNN-based baseline.  Furthermore, \WARP{} demonstrates a 3x speedup over the state-of-the-art ColBERTv2/PLAID system, as illustrated in Figure~\ref{fig:comp_lotte_pooled_test}.

After reviewing prior work on efficient neural IR in \Cref{chapter:related_work}, we analyze the latency bottlenecks in the ColBERT and XTR retrieval frameworks in \Cref{chapter:latency_breakdown}, identifying key areas for optimization. These findings form the foundation for our work on \WARP{}, which we introduce and describe in detail in \Cref{chapter:warp_engine}. In \Cref{chapter:experiments}, we evaluate \WARP{}'s end-to-end latency and scalability using the BEIR \cite{beir} and LoTTE \cite{colbert2} benchmarks. Finally, we compare our implementation to existing state-of-the-art engines.

\section{Related Work}

\label{chapter:related_work}

Dense retrieval models can be broadly categorized into single-vector and multi-vector approaches. Single-vector methods, exemplified by ANCE \cite{ance} and STAR/ADORE \cite{star}, encode a passage into a single dense vector \cite{singlevector}. While these techniques offer computational efficiency, their inherent limitation of representing complex documents with a single vector has been shown to constrain the model's ability to capture intricate information structures \cite{colbert}.

To address such limitations, ColBERT \cite{colbert} introduces a multi-vector paradigm. Here, both queries and documents are independently encoded as multiple embeddings, allowing for a richer representation of document content and query intent.
The multi-vector approach is further refined in ColBERTv2 \cite{colbert2}, which improves supervision and incorporates residual compression to reduce the space requirements associated with storing multiple vectors per indexed document. Building upon these innovations, PLAID \cite{plaid} substantially accelerates ColBERTv2 by efficiently pruning non-relevant passages using the residual representation and by employing optimized C++ kernels for decompression and scoring. EMVB \cite{emvb} further optimizes PLAID's memory usage and \emph{single-threaded} query latency using product quantization \cite{product_quantization} and SIMD instructions.

Separately, COIL \cite{coil} incorporates insights from conventional retrieval systems \cite{bm25} by constraining token interactions to lexical matches between queries and documents. SPLATE \cite{splate} translates the embeddings produced by ColBERTv2 style pipelines to a sparse vocabulary, allowing the candidate generation step to be performed using traditional \emph{sparse} retrieval techniques. CITADEL \cite{citadel} introduces conditional token interaction through dynamic lexical routing, selectively considering tokens for relevance estimation. While CITADEL significantly reduces GPU execution time, it falls short of PLAID's CPU performance at comparable retrieval quality.

The ConteXtualized Token Retriever (XTR) \cite{xtr} represents a notable conceptual advancement in dense retrieval. XTR simplifies the scoring process and eliminates the gathering stage entirely, \emph{theoretically} enhancing retrieval efficiency. However, its current end-to-end latency limits its applicability in production environments, where even minor increases in query response time can degrade user experience and negatively affect revenue \cite{revenue}.

\section{Latency of Current Neural Retrievers}

\label{chapter:latency_breakdown}

We start by analyzing two state-of-the-art multi-vector retrieval methods to identify their bottlenecks, providing the foundation for our work on \WARP{}. We evaluate the latency of PLAID and XTR across various configurations and datasets: BEIR NFCorpus, LoTTE Lifestyle, and LoTTE Pooled. In XTR, token retrieval emerges as a fundamental bottleneck: the need to retrieve a large number of candidates from the ANN backend significantly impacts performance. PLAID, while generally more efficient, faces challenges in its decompression stage. Query encoding emerges as a shared limitation for both engines, particularly on smaller datasets. These insights inform the design of \WARP{}, which we introduce in~\Cref{chapter:warp_engine}.

\subsection{ColBERTv2/PLAID}

\begin{figure}[t]
    \centering
    \includegraphics[width=0.45\textwidth]{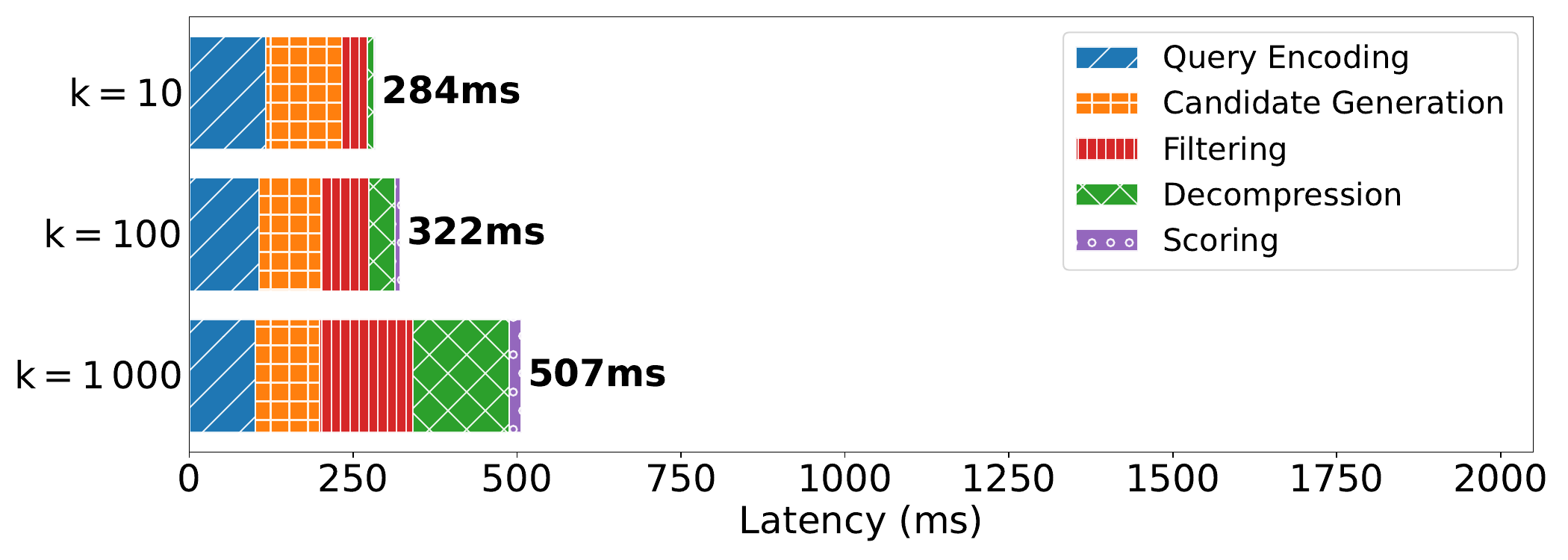}
    \caption{Breakdown of ColBERTv2/PLAID's avg. latency for varying $k$ on LoTTE Pooled (Dev Set)}
    \label{fig:comp_lotte_pooled_test_v}
\end{figure}

As shown in \Cref{fig:comp_lotte_pooled_test_v}, we evaluate PLAID's performance using its optimized implementation \cite{colbert_code} and the ColBERTv2 checkpoint from Hugging Face \cite{colbert_huggingface}. We configure PLAID's hyperparameters similar to the original paper \cite{plaid}. Consistent with prior work \cite{emvb}, we observe single threaded CPU latency exceeding 500ms on LoTTE Pooled. Furthermore, we find that the decompression stage remains rather constant for fixed $k$ across all datasets, consuming approximately 150-200ms for $k = 1\,000$. Notably, for smaller datasets like BEIR NFCorpus and large $k$ values, this stage constitutes a significant portion of the overall query latency. %
Thus, the decompression stage emerges as a critical bottleneck for small datasets. As anticipated, the filtering stage's execution time is proportional to the number of candidates, increasing for larger $k$ values and bigger datasets. In contrast, candidate generation constitutes a fixed cost based on the number of centroids. Interestingly, the scoring stage appears to have a negligible impact on ColBERTv2/PLAID's overall latency across all measurements.

\subsection{XTR/ScaNN}

To enable benchmarking of the XTR framework, we develop a Python library based on Google DeepMind's published code \cite{xtr_code}. We provide the library’s code and scripts to reproduce the benchmarks on GitHub\footnote{ \url{https://github.com/jlscheerer/xtr-eval}}.
Unless otherwise specified, all benchmarks utilize the XTR \texttt{BASE\_EN} transformer model for encoding. This model was published and is available on Hugging Face \cite{xtr_base_en_huggingface}. In accordance with the paper \cite{xtr}, we evaluate the implementation for k'$\ = 1\,000$ and k'$\ = 40\,000$.

\begin{figure}[t]
    \centering
    \begin{subfigure}[t]{0.45\textwidth}
        \centering
        \includegraphics[width=\textwidth]{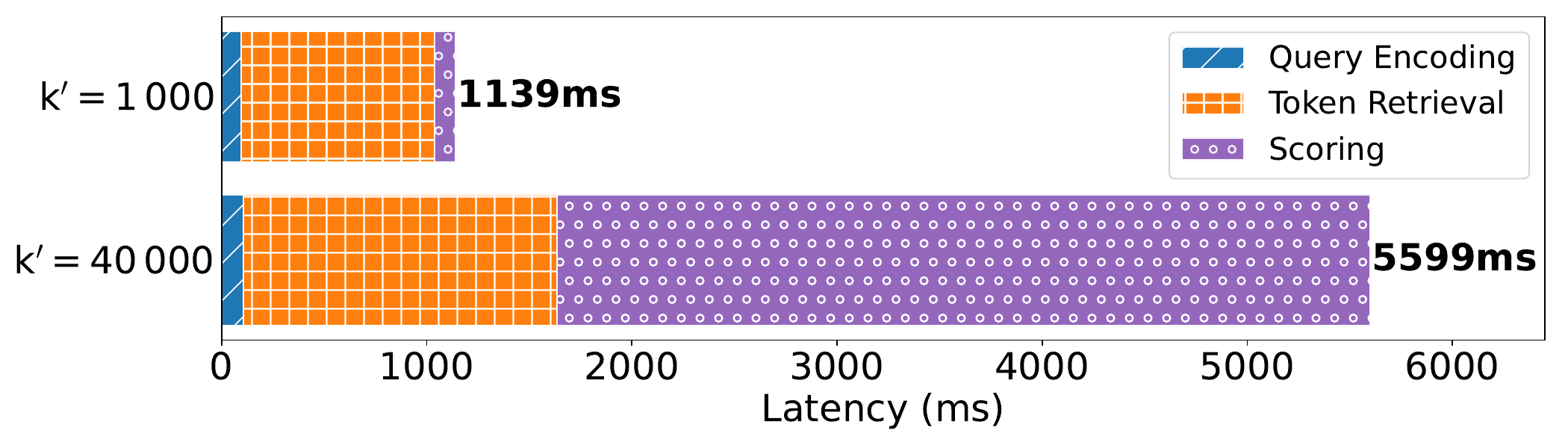}
        \caption{Latency breakdown of the reference implementation.}
        \label{fig:unopt_lotte_pooled_dev}
        \vspace{0.5em}
    \end{subfigure}
    \begin{subfigure}[t]{0.45\textwidth}
        \centering
        \includegraphics[width=\textwidth]{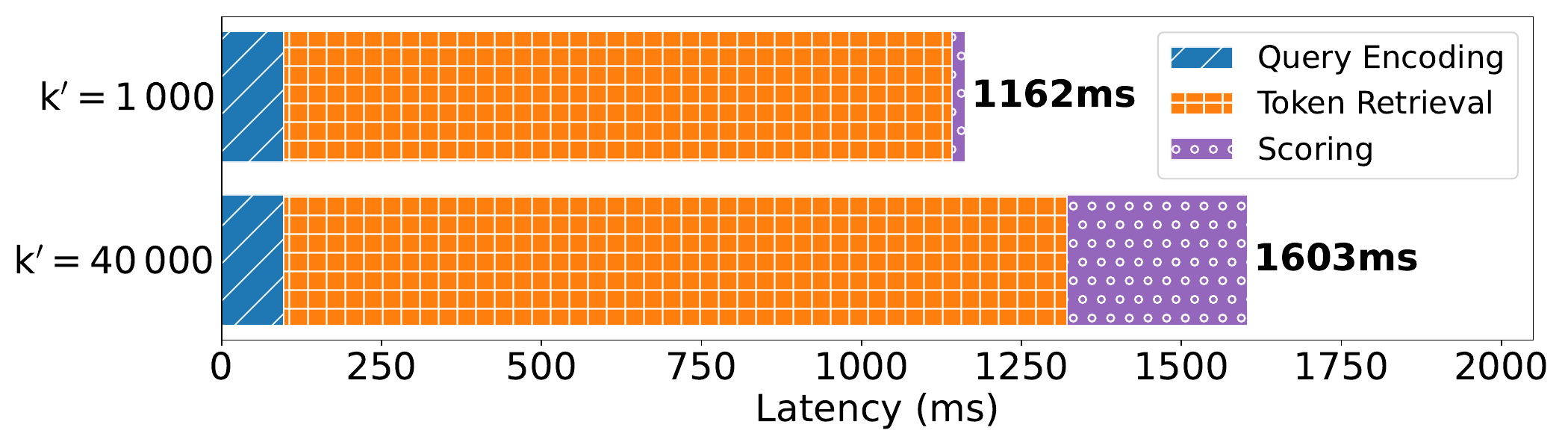}
        \caption{Latency breakdown of the optimized implementation.}
        \label{fig:comp_lotte_pooled_dev}
    \end{subfigure}
    \caption{Breakdown of XTR$_\text{base}$/ScaNN on LoTTE Pooled.}
    \label{fig:xtr_lotte_comparison}
\end{figure}

As seen in \Cref{fig:unopt_lotte_pooled_dev}, the scoring stage constitutes a significant bottleneck in the end-to-end latency of the XTR framework, particularly when dealing with large values of $k'$. We argue that this performance bottleneck is largely attributed to an \emph{unoptimized} implementation in the released code, which relies on native Python data structures and manual iteration, introducing substantial overhead, especially for large numbers of token embeddings. We refactor this implementation to leverage optimized data structures and vectorized operations. This helps us uncover hidden performance inefficiencies and establish a baseline for further optimization. Our improved XTR implementation is publicly available in the corresponding GitHub repository\footnote{\url{https://github.com/jlscheerer/xtr-eval}}.

We present the evaluation of our optimized implementation in \Cref{fig:comp_lotte_pooled_dev}. Notably, the optimized implementation's end-to-end latency is significantly lower than that of the reference implementation ranging from an end-to-end 3.5x speedup on LoTTE pooled to a 6.3x speedup on LoTTE Lifestyle for $k = 1\,000$. This latency reduction is owed in large parts to a more efficient \emph{scoring} implementation -- a 14x speedup on LoTTE Pooled. In particular, this reveals token retrieval as the fundamental bottleneck of the XTR framework. Although the optimized scoring stage accounts for a small fraction of the overall end-to-end latency, it is still slow in absolute terms, ranging from 33ms to 281ms for $k = 1\,000$ on BEIR NFCorpus and LoTTE Pooled, respectively.

\section{WARP}

\label{chapter:warp_engine}

\WARP{} optimizes retrieval for the refined late interaction architecture introduced in XTR. Seeking to find the best of both the XTR and PLAID worlds, \textbf{\WARP{} introduces the
novel WARP$_\text{SELECT}$ algorithm for candidate generation, which effectively avoids gathering token-level representations, and proposes an optimized two-stage reduction for faster scoring via a dedicated C++ kernel combined with implicit decompression.} \WARP{} also uses specialized inference runtimes for faster query encoding.

As in XTR, queries and documents are encoded \emph{independently} into embeddings at the token-level using a fine-tuned T5 transformer \cite{t5transformer}. To scale to large datasets, document representations are computed in advance and constitute \WARP{}'s index. The similarity between a query $q$ and document $d$ is modeled using XTR's adaptation of ColBERT's summation of MaxSim operations\footnote{Note that we omit the normalization via $\frac{1}{Z}$, as we are only interested in the relative ranking between documents and the normalization constant is identical for any retrieved document.}:
\begin{equation}
    S_{d,q} = \sum_{i=1}^n \max_{1 \leq j \leq m} [\mathbf{\hat{A}}_{i,j}q_i^Td_j + (1 - \mathbf{\hat{A}}_{i,j})m_i]
    \label{eq:xtr_relevance_score_inference}
\end{equation}
where $q$ and $d$ are the matrix representations of the query and passage embeddings respectively, $m_i$ denotes the missing similarity estimate for $q_i$, and $\mathbf{\hat{A}}$ describes XTR's alignment matrix. In particular, $\mathbf{\hat{A}}_{i,j} = \mathbbm{1}_{[j \in \text{top-}k_j'(\mathbf{d}_{i,j'})]}$ captures whether a document token embedding of a candidate passage was retrieved for a specific query token $q_i$ as part of the token retrieval stage. We refer to \cite{colbert, xtr} for an intuition behind this choice of scoring function.

\subsection{Index Construction}
\label{sub:warp_index_construction}

Akin to ColBERTv2 \cite{colbert2}, \WARP{}'s compression strategy involves applying $k$-means clustering to the produced document token embeddings. As in ColBERTv2, we find that using a sample of all passages proportional to the square root of the collection size to generate this clustering performs well in practice. 
After having clustered the sample of passages, all token-level vectors are encoded and stored as quantized residual vectors to their nearest cluster centroid. Each dimension of the quantized residual vector is a $b$-bit encoding of the delta between the centroid and the original uncompressed vector.  In particular, these deltas are stored as a sequence of $128 \cdot \frac{b}{8}$ $8$-bit values, wherein $128$ represents the transformer's token embedding dimension. Typically, we set $b = 4$, i.e., compress each dimension of the residual into a single nibble, for an $8x$ compression\footnote{As compared to an uncompressed $32$-bit floating point number per dimension.}. In this case, each compressed $8$-bit value stores 2 indices in the range $[0, 2^b)$. Instead of quantizing the residuals uniformly, \WARP{} uses quantiles derived from the empirical distribution to determine bucket boundaries and the corresponding representative values. This process allows \WARP{} to allocate more quantization levels to densely populated regions of the data distribution, thereby minimizing the overall quantization error for residual compression.

\subsection{Retrieval}

\begin{figure}[t]
    \centering
    \resizebox{0.45\textwidth}{!}{
        \input{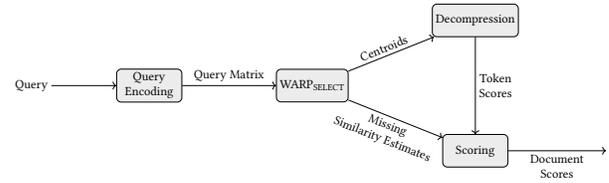}
    }
    \caption[\WARP{}'s Inference Pipeline]{\WARP{} Retrieval consisting of query encoding, WARP$_\text{SELECT}$, decompression, and scoring. Notably, centroid selection is combined with the computation of missing similarity estimates in WARP$_\text{SELECT}$.}
    \label{fig:warp_pipeline}
\end{figure}

Extending PLAID, the retrieval process in the \WARP{} engine is divided into four distinct steps: query encoding, candidate generation, decompression, and scoring. \Cref{fig:warp_pipeline} illustrates the retrieval process in \WARP{}. The process starts by encoding the query text into $q$, a $(\text{query\_maxlen}, 128)$-dimensional tensor, using the underlying Transformer model.\footnote{We set $\text{query\_maxlen} = 32$ in accordance with the XTR paper.} Next, the most similar $n_{\text{probe}}$ centroids are identified for each of the $\text{query\_maxlen}$ query token embeddings.

Subsequently, \WARP{} identifies all document token embeddings belonging to the clusters of the selected centroids and computes their \emph{individual} relevance score. Computing this score involves decompressing residuals of the identified document token embeddings and calculating their cosine similarity with the relevant query token embedding. Finally, \WARP{} implicitly constructs an $n_\text{candidates} \times \text{query\_maxlen}$ score matrix $S$\footnote{Note that our optimized implementation does not physically construct this matrix.}, where each entry $S_{d_i,q_j}$ contains the maximum retrieved score for the $i$-th candidate passage $d_i$ and  the $j$-th query token embedding $q_j$:\begin{equation*}
    \max_{1 \leq j \leq m} \mathbf{\hat{A}}_{i,j}q_i^Td_j
\end{equation*} Matrix entries not populated during token retrieval, i.e., $\mathbf{\hat{A}}_{i,j} = 0$, are imputed with a missing similarity estimate, as postulated in the XTR framework. To compute the relevance score of a document $d_i$, the cumulative score over all query tokens is computed: $\sum_j S_{d_i,q_j}$. To produce the ordered set of passages, the set of scores is sorted and the top $k$ highest scoring passages are returned. 

\footnotetext{For XTR baselines, 'candidate generation' refers to the token retrieval stage.}

\subsection{WARP$_\text{SELECT}$}

In contrast to ColBERT, which populates the entire score matrix for the items retrieved, XTR only populates the score matrix with scores computed as part of the token retrieval stage. To account for the contribution of any missing tokens, XTR relies on \emph{missing similarity imputation}, in which they set any missing similarity of the query token for a specific document as the lowest score obtained as part of the token retrieval stage. The authors argue that this approach is justified as it constitutes a \emph{natural} upper bound for the true relevance score. In the case of \WARP{}, this bound is no longer guaranteed to hold.\footnote{Even in XTR’s case, the bound is only approximate, as ScaNN does not guarantee exact nearest neighbors.}

Instead, \WARP{} defines a novel strategy for missing similarity imputation based on cumulative cluster sizes, WARP$_\text{SELECT}$. Given the query embedding matrix $q$ and the list of centroids $C$ (\Cref{sub:warp_index_construction}), \WARP{} computes the token-level query-centroid relevance scores. As both the query embedding vectors and the set of centroids are normalized, the cosine similarity scores $S_{c,q}$ can be computed efficiently as a matrix product:
\begin{equation*}
    S_{c,q} = C \cdot q^T
\end{equation*}
Once these relevance scores have been computed, \WARP{} identifies the $n_{\text{probe}}$ centroids with the largest similarity scores for decompression, as part of candidate generation.

\textbf{Using these query--centroid similarity scores, WARP$_\text{SELECT}$ folds the estimation of missing similarity scores into candidate generation. Specifically, for each query token $q_i$, it sets $m_i$ from \Cref{eq:xtr_relevance_score_inference} as the first element in the sorted list of centroid scores for which the cumulative cluster size exceeds a threshold $t'$.} This method is particularly attractive as all the centroid scores have already been computed and sufficiently sorted as part of candidate generation, so the cost of computing missing similarity imputation with this method is negligible.

We find that $t'$ is easy to configure (see \Cref{sec:hyperparameters}) without compromising retrieval quality or efficiency. This represents a significant improvement over the missing similarity estimate used in the XTR reference implementation, where the estimate is \emph{inherently} tied to the number of retrieved tokens. Intuitively, increasing $k'$ in XTR may only help refine the missing similarity estimate, but not necessarily increase the density of the score matrix.\footnote{This is because tokens retrieved with a larger $k’$ are often from new documents and, therefore, do not refine the scores of already retrieved ones.}

\subsection{Decompression}

The input for the decompression phase is the set of $n_\text{probe}$ centroid indices for each of the \texttt{query\_maxlen} query tokens. Its goal is to calculate relevance scores between each query token and the embeddings within the identified clusters.
For a query token $q_i$, let $c_{i,j}$, where $j \in [n_\text{probe}]$, be the set of centroid indices identified during candidate generation. Let ${r_{i,j,k}}$ be the set of residuals associated with cluster $c_{i,j}$. The decompression step computes:
\begin{equation}
s_{i,j,k} = \text{decompress}(C[c_{i,j}], r_{i,j,k}) \times q_i^T\ \forall\ i,j,k
\end{equation}

The decompress function converts residuals from their compact representation into uncompressed vectors.
Each residual $r_{i,j,k}$ is composed of 128 indices, each $b$ bits wide. These indices reference values in the bucket weights vector $\omega \in \mathbb{R}^{2^b}$ and are used to offset the centroid $C[c_{i,j}]$. The $\text{decompress}$ function is defined as:
\begin{equation}
\text{decompress}(C[c_{i,j}], r_{i,j,k}) = C[c_{i,j}] + \sum_{d=1}^{128}\bm{e}_d \cdot \omega[(r_{i,j,k})_{d}]
\end{equation}
Here, $\bm{e}_d$ is the unit vector for dimension $d$, and $\omega[(r_{i,j,k})_{d}]$ is the weight value at index $(r_{i,j,k})_{d}$ for dimension $d$. In other words, the indices are used to look up specific entries in $\omega$ for each dimension independently, adjusting the centroid accordingly.

\textbf{\WARP{} avoids \emph{explicitly} decompressing residuals, as PLAID does, by leveraging the observation that the scoring function decomposes between centroids and residuals. As a result, \WARP{} reuses the query-centroid relevance scores $S_{c,q}$, computed as part of candidate generation.} That is, observe that:
\begin{equation}
\begin{split}
\label{eq:warp_selective_sum}
	s_{i,j,k} &= \text{decompress}(C[c_{i,j}], r_{i,j,k}) \times q_i^T \\
	&= (C[c_{i,j}] \times q_i^T) + (\sum_{d=1}^{128} \omega[(r_{i,j,k})_{d}] q_{i,d})
\end{split}
\end{equation}

To accelerate decompression, \WARP{} computes $\upsilon = \hat{q} \times \hat{\omega}$, wherein $\hat{q} \in \mathbb{R}^{\texttt{query\_maxlen} \times 128 \times 1}$ represents the query matrix that has been \emph{unsqueezed} along the last dimension, and $\hat{\omega} \in \mathbb{R}^{1 \times 2^b}$ denotes the vector of bucket weights that has been \emph{unsqueezed} along the first dimension. 

With these definitions, \WARP{} can decompress and score candidate tokens via:
\begin{equation}
\begin{split}
	s_{i,j,k} &= S_{c_j,q_i} + \sum_{d=1}^{128}(\omega \cdot q_{i,d})[(r_{i,j,k})_{d}] \\
	&= S_{c_j,q_i} + \sum_{d = 1}^{128}\upsilon_{i,d}[(r_{i,j,k})_{d}]
\end{split}
\end{equation}

Note that candidate scoring can now be implemented as a simple \emph{selective sum}. As the bucket weights are shared among centroids and the query-centroid relevance scores have already been computed during candidate generation, \WARP{} can decompress and score arbitrarily many clusters using $O(1)$ multiplications. \textbf{This refined scoring function is far more efficient than the one outlined in PLAID,\footnote{PLAID cannot adopt this approach directly, as it \emph{normalizes} the vectors after decompression. Empirically, we find that this normalization step has little effect on the final embeddings as the residuals are already normalized prior to quantization.} as it never computes the decompressed embeddings explicitly and instead directly emits the resulting candidate scores.} We provide an efficient implementation of the selective sum of \Cref{eq:warp_selective_sum} and realize unpacking of the residual representation using low-complexity bitwise operations as part of \WARP{}'s C++ kernel for decompression.

\subsection{Scoring}
\label{subsection:warp_scoring}

At the end of the decompression phase, we have $\texttt{query\_maxlen} \times n_\text{probe}$ strides of decompressed candidate document \emph{token-level scores} and their corresponding document identifiers. Scoring combines these scores with the missing similarity estimates, computed during candidate generation, to produce \emph{document-level scores}. This process corresponds to constructing the score matrix and taking the row-wise sum.

Explicitly constructing the score matrix, as in the reference XTR implementation, introduces a significant bottleneck, particularly for large values of $n_\text{probe}$. To address this, \WARP{} efficiently aggregates token-level scores using a two-stage reduction process:
\begin{itemize}
	\item \textbf{Token-level reduction} For each query token, reduce the corresponding set of $n_\text{probe}$ strides using the \texttt{max} operator. This step \emph{implicitly} fills the score matrix with the maximum per-token score for each document. As a single cluster can contain multiple document token embeddings originating from the same document, \WARP{} performs \emph{inner-cluster} \texttt{max}-reduction directly during the decompression phase.
	\item \textbf{Document-level reduction} Reduce the resulting strides into document-level scores using a \texttt{sum} aggregation. It is essential to handle missing values properly at this stage -- any missing per-token score must be replaced by the corresponding missing similarity estimate, ensuring compliance with the XTR scoring function described in \Cref{eq:xtr_relevance_score_inference}. This reduction step corresponds to the row-wise summation of the score matrix.
	
\end{itemize}

After performing both reduction phases, the final stride contains the document-level scores and the corresponding identifiers for all candidate documents. To retrieve the result set, we perform heap select to obtain the top-$k$ documents, similar to its use in the candidate generation phase.

Formally, we consider a stride $S$ to be a list of key-value pairs:
\begin{equation*}
    S = \{(k_i, v_i)\};\ \text{K}(S) = \{k_i\ |\ (k_i, v_i) \in S\};\ \text{V}(S) = \{v_i\ |\ (k_i, v_i) \in S\}\ 
\end{equation*}
Thus, strides implicitly define a partial function $f_S : K \rightharpoonup V(S)$:
\begin{equation*}
    f_S(k) = \begin{cases}
        v_i  & \text{if }\exists v_i.\ (k, v_i) \in S \\
        \bot & \text{otherwise}
    \end{cases}
\end{equation*}
We define a reduction as a combination of two strides $S_1$ and $S_2$ using a binary function $r$ into a single stride by applying $r$ to values of matching keys:
\begin{equation*}
    \text{reduce}(r, S_1, S_2) = \{(k, r(f_{S_1}(k), f_{S_2}(k)))\ |\ k \in K(S_1) \cup K(S_2)\}
\end{equation*}

With these definitions, token-level reduction can be written as:
\begin{equation}
    r_\text{tok}(v_1, v_2) = \begin{cases}
        \max(v_1, v_2) & \text{if } v_1 \neq \bot \land v_2 \neq \bot \\
        v_1 & \text{if } v_1 \neq \bot \land v_2 = \bot \\
        v_2 & \text{otherwise }
    \end{cases}
\end{equation}

Defining document-level reduction is slightly more complex as it involves incorporating the {corresponding missing similarity estimates $m$. After token-level reduction each of the \texttt{query\_maxlen} strides $S_1, \dots, S_{\texttt{query\_maxlen}}$ \emph{covers} scores for a single query token $q_i$. We set $S_{i,i} = S_i$ and define:
\begin{equation}
    S_{i,j} = \text{reduce}(r_{\text{doc},(i,k,j)}, S_{i,k}, S_{k + 1, j})
    \label{eq:merge_strides}
\end{equation}
for any choice of $i \leq k < j$, wherein $r_{\text{doc},(i,k,j)}$ merges two successive, non-overlapping strides $S_{i,k}$ and $S_{k + 1, j}$. The resulting stride, $S_{i,j}$, now covers scores for query tokens $q_i, \dots, q_j$. Defining $r_{\text{doc},(i,k,j)}$ is relatively straightforward:
\begin{equation}
    r_{\text{doc},(i,k,j)}(v_1, v_2) = \begin{cases}
        v_1 + v_2 & \text{if } v_1 \neq \bot \land v_2 \neq \bot \\
        v_1 + (\sum_{t = k + 1}^{j} m_t) & \text{if } v_1 \neq \bot \land v_2 = \bot \\
        (\sum_{t = i}^{k} m_t) + v_2 & \text{otherwise }
    \end{cases}
    \label{eq:doc_level_reduce}
\end{equation}

It is easy to verify that $S_{i,j}$ is well-defined, i.e., independent of the choice of $k$. The result of document-level reduction is $S_{1,\texttt{query\_maxlen}}$ and can be obtained by recursively applying \Cref{eq:merge_strides} to strides of increasing size.

\WARP{}'s two-stage reduction process, along with the final sorting step, is illustrated in \Cref{fig:two_stage_reduce_aggregate}. In the token-level reduction stage, strides are merged by selecting the maximum value for matching keys. In the document-level reduction stage, values for matching keys are summed, with missing values being substituted by the corresponding missing similarity estimates.

\begin{figure}[t]
    \centering
    \resizebox{0.45\textwidth}{!}{
        \input{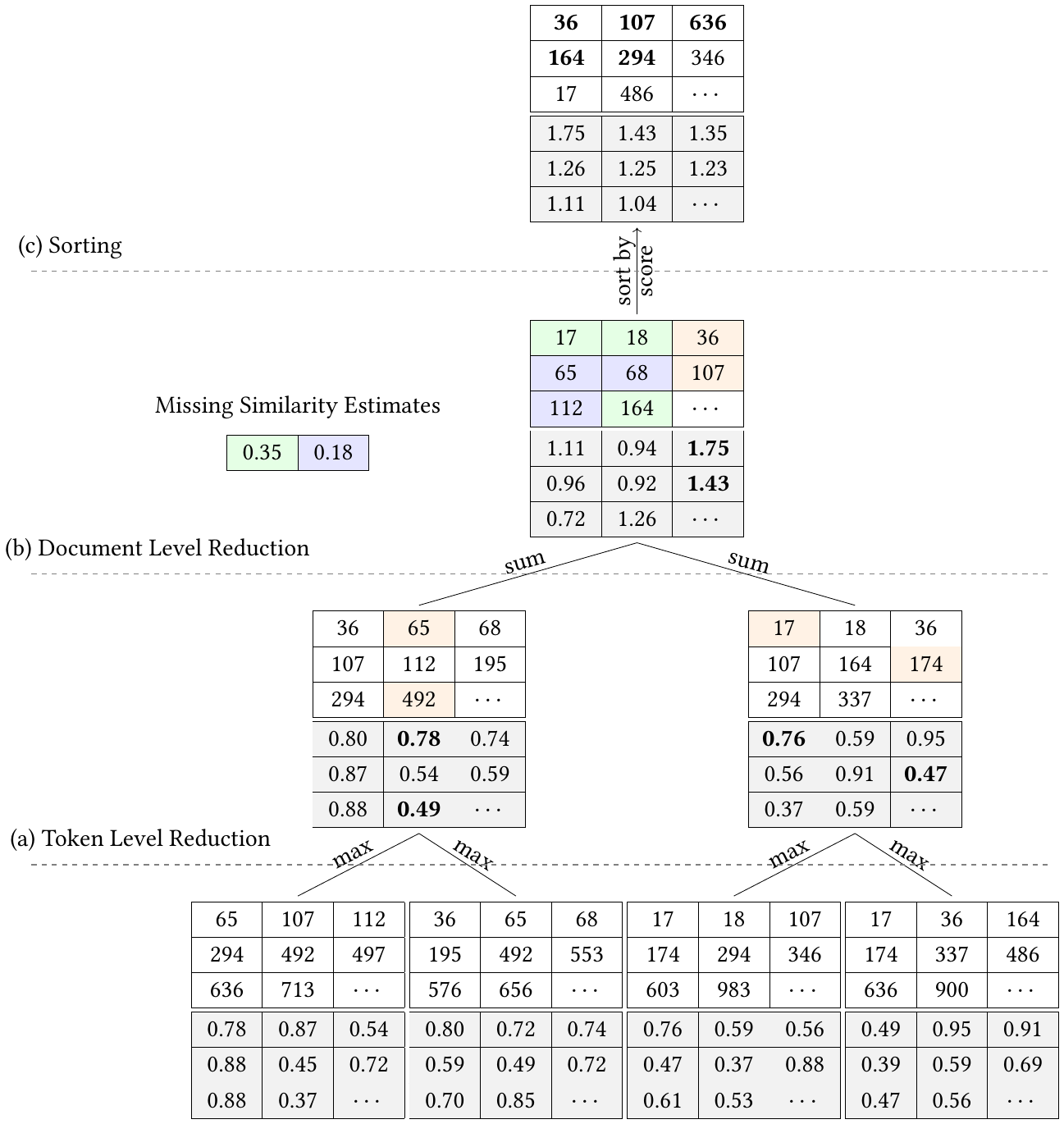}
    }
    \caption{\WARP{}'s scoring phase: (a) In token-level reduction, strides are max-reduced. (b) In document-level reduction, values are summed, accounting for missing similarity estimates. (c) Scores are sorted, yielding the top-$k$ results.}
    \label{fig:two_stage_reduce_aggregate}
\end{figure}

In our implementation, we conceptually construct a binary tree of the required merges and alternate between using two scratch buffers to avoid additional memory allocations. We realize \Cref{eq:doc_level_reduce} using a prefix sum, which involves precalculating running totals to eliminate the need to compute sums explicitly later on.

\subsection{Hyperparameters}
\label{sec:hyperparameters}

In this section, we analyze the effects of \WARP{}'s three primary hyperparameters, namely: 
\begin{itemize}
    \item $n_\text{probe}$ -- the \#clusters to decompress per query token
    \item $t'$ -- the threshold on the cluster size used for WARP$_\text{SELECT}$
    \item $b$ -- the number of bits per dimension of a residual vector
\end{itemize}

To study the effects of $n_\text{probe}$ and $t'$, we analyze the normalized Recall@100\footnote{The normalized Recall@k is calculated by dividing Recall@k by the dataset's maximum, effectively scaling values between 0 and 1 to ensure comparability across datasets.} as a function of $t'$ for $n_\text{probe} \in \{1,2,4,8,16,32,64\}$ across four development datasets of increasing size: BEIR NFCorpus, BEIR Quora, LoTTE Lifestyle, and LoTTE Pooled. For further details on the datasets, please refer to \Cref{table:datasets_info}. \Cref{fig:hyper_nprobe} visualizes the results of our analysis. We observe a consistent pattern across all evaluated datasets, namely substantial improvements as $n_\text{probe}$ increases from 1 to 16 (i.e., 1, 2, 4, 8, 16), followed by only marginal gains in Recall@100 beyond that. A notable exception is BEIR NFCorpus, where we still observe significant improvement when increasing from $n_\text{probe} = 16$ to $n_\text{probe} = 32$. We hypothesize that this is due to the small number of embeddings per cluster in NFCorpus, limiting the number of scores available for aggregation. Consequently, we conclude that setting $n_\text{probe} = 32$ strikes a good balance between end-to-end latency and retrieval quality.

In general, we find that \WARP{} is highly robust to variations in $t'$. However, smaller datasets, such as NFCorpus, appear to benefit from a smaller $t'$, while larger datasets perform better with a larger $t'$. Empirically, we find that setting $t'$ proportional to the square root of the dataset size consistently yields strong results across all datasets. Moreover, increasing $t'$ beyond a certain point no longer improves recall, leading us to bound $t'$ by a maximum value, $t'_\text{max}$.

\begin{figure}[t]
    \begin{subfigure}[t]{0.235\textwidth}
        \centering
        \includegraphics[width=\linewidth]{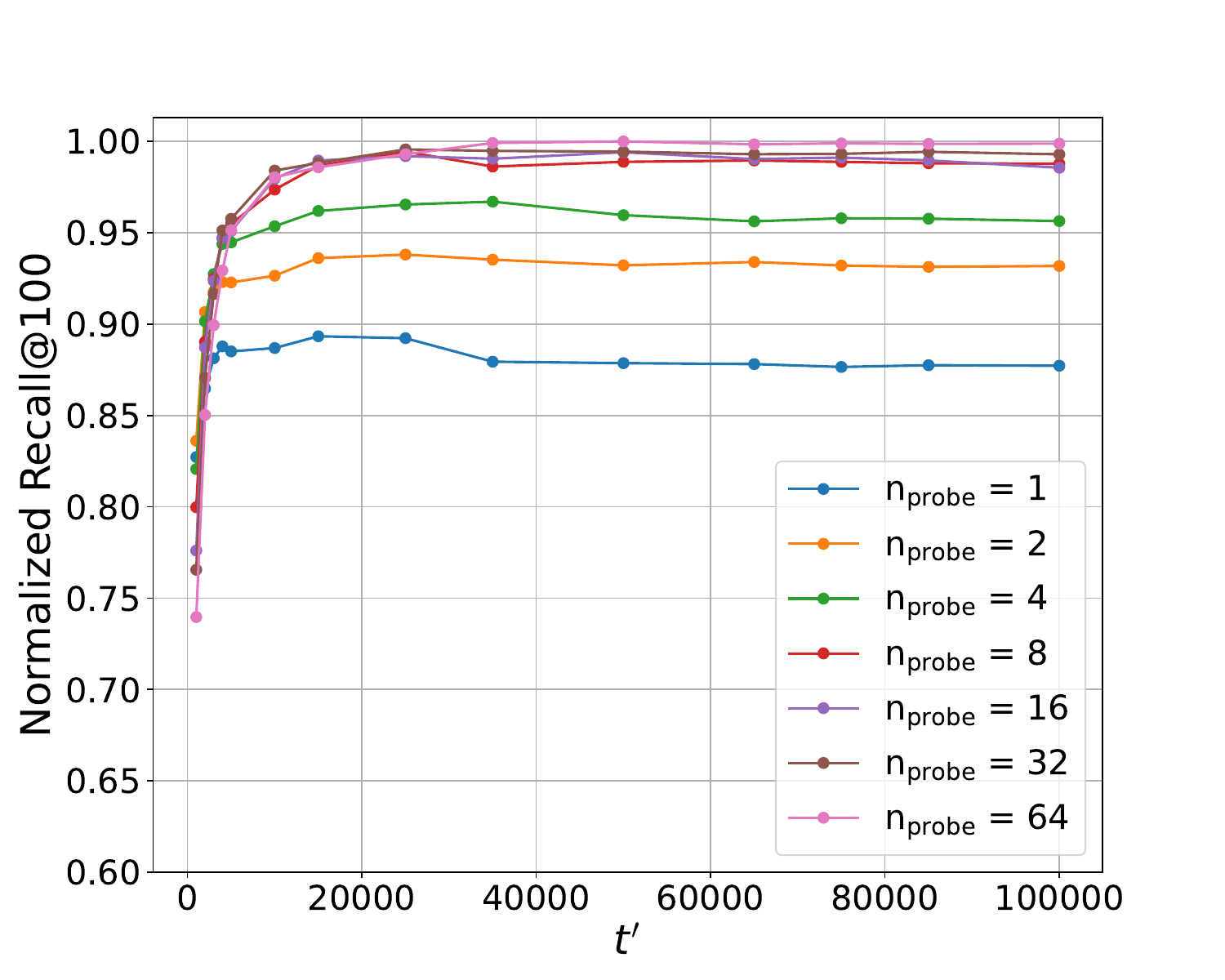}
        \caption{LoTTE Science (Dev Set)}
    \end{subfigure}
    \hfill
    \begin{subfigure}[t]{0.235\textwidth}
        \centering
        \includegraphics[width=\linewidth]{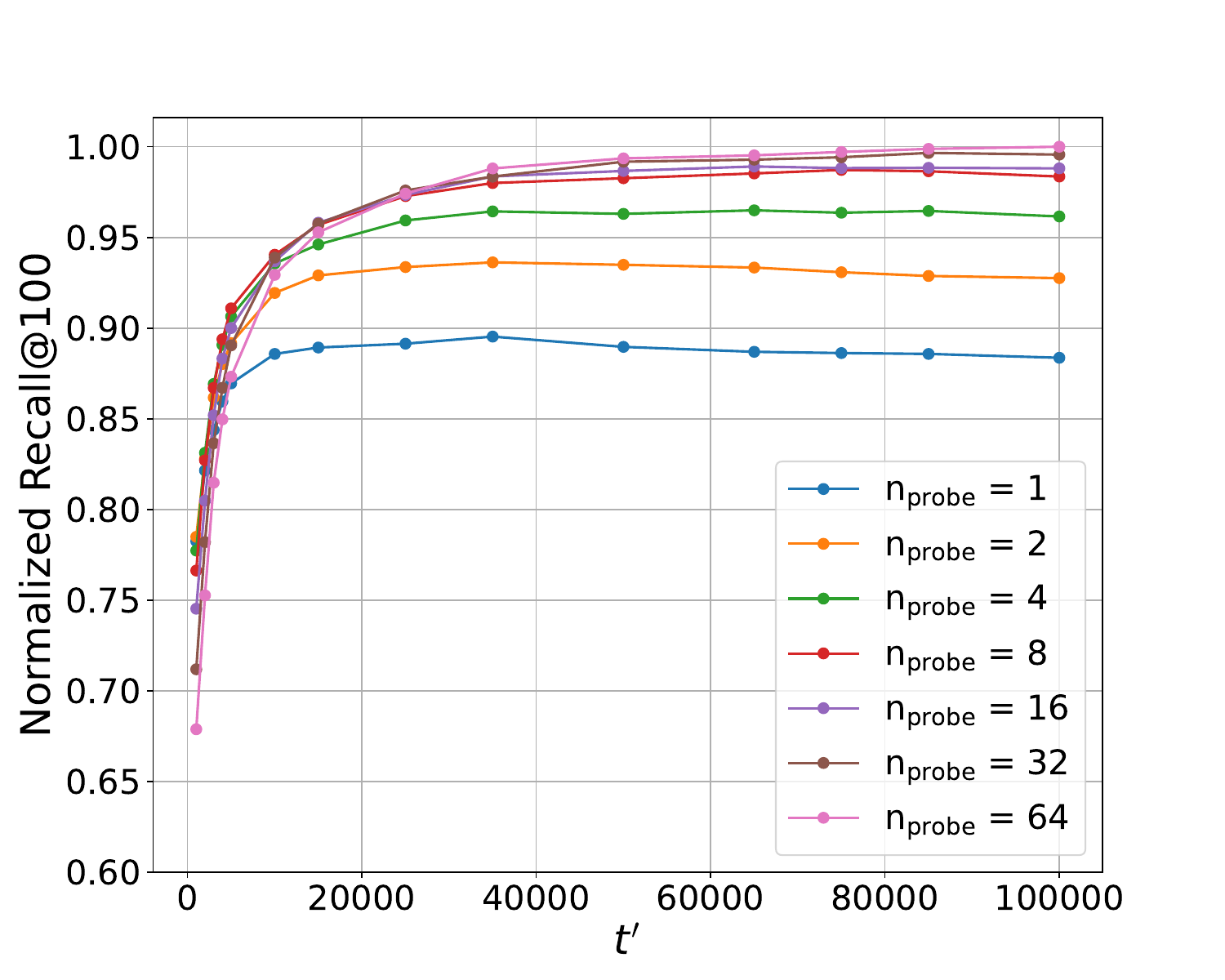}
        \caption{LoTTE Pooled (Dev Set)}
    \end{subfigure}
    \caption{nRecall@100 as a function of $t'$ and $n_\text{probe}$}
    \label{fig:hyper_nprobe}
\end{figure}

Next, we aim to quantify the effect of $b$ on the retrieval quality of \WARP{}, as shown in \Cref{fig:hyper_b}. To do this, we compute the nRecall@k for $n_\text{probe} = 32$ and $k \in \{10, 100\}$ using two datasets: LoTTE Science and LoTTE Pooled. 

Our results show a significant improvement in retrieval performance when increasing $b$ from 2 to 4, particularly for smaller values of $k$. For larger values of $k$, the difference in performance diminishes, particularly for the LoTTE Pooled dataset.

\begin{figure}[t]
    \begin{subfigure}[t]{0.235\textwidth}
        \centering
        \includegraphics[width=\linewidth]{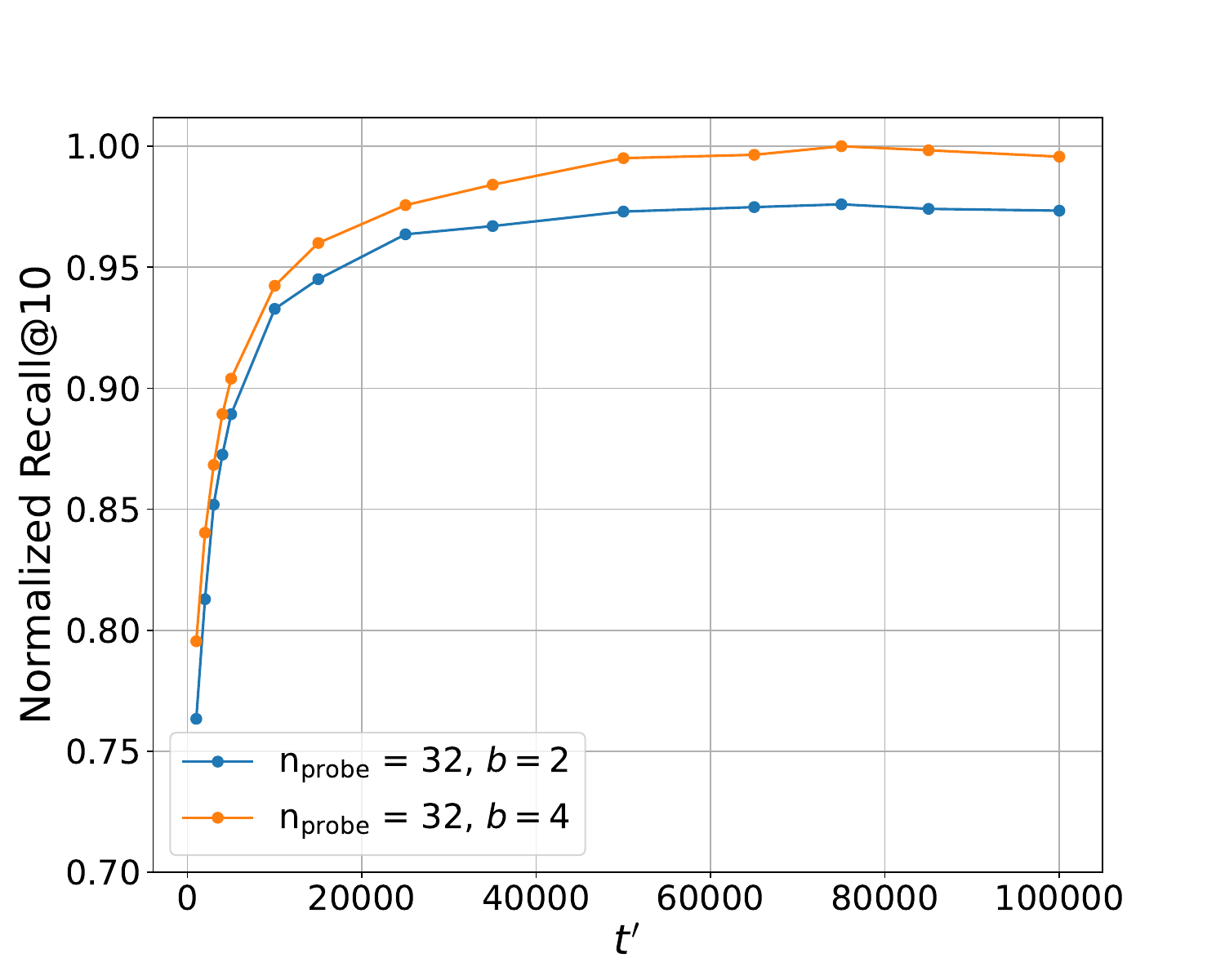}
        \caption{LoTTE Pooled (Dev Set), nRecall@10}
    \end{subfigure}
    \hfill
    \begin{subfigure}[t]{0.235\textwidth}
        \centering
        \includegraphics[width=\linewidth]{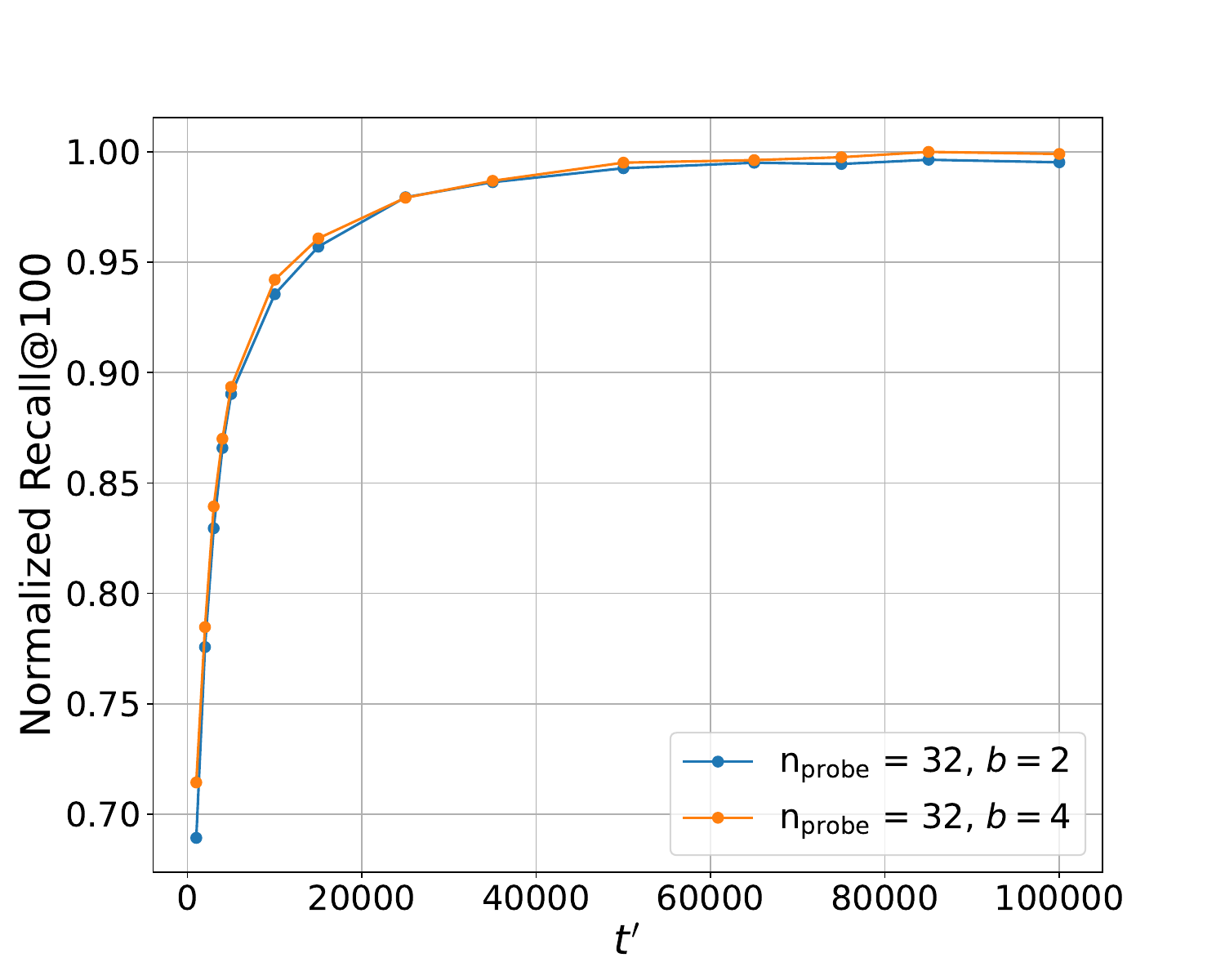}
        \caption{LoTTE Pooled (Dev Set), nRecall@100}
    \end{subfigure}
    \caption{nRecall@k as a function of $t'$ and $b$}
       \label{fig:hyper_b}
\end{figure}

\section{Evaluation}

\label{chapter:experiments}

\begin{table}[t]
    \centering
    \resizebox{\columnwidth}{!}{%
    \begin{tabular}{@{}llrrcrr@{}}
        \toprule
        \multicolumn{2}{c}{Dataset} & \multicolumn{2}{c}{Dev} & & \multicolumn{2}{c}{Test} \\ 
        \cmidrule(r){3-4} \cmidrule(l){6-7}
         &  & \#Queries & \#Passages & & \#Queries & \#Passages \\ 
        \midrule
        \multirow{6}{*}{\textbf{BeIR}~\cite{beir}}
        & NFCORPUS & 324 & 3.6K & & 323 & 3.6K \\
        & SciFact & -- & -- & & 300 & 5.2K \\
        & SCIDOCS & -- & -- & & 1,000 & 25.7K \\
        & Quora & 5,000 & 522.9K & & 10,000 & 522.9K \\
        & FiQA-2018 & 500 & 57.6K & & 648 & 57.6K \\
        & Touché-2020 & -- & -- & & 49 & 382.5K \\
        \midrule
        \multirow{6}{*}{\textbf{LoTTE}~\cite{colbert2}}
        & Lifestyle & 417 & 268.9K & & 661 & 119.5K \\
        & Recreation & 563 & 263.0K & & 924 & 167.0K \\
        & Writing & 497 & 277.1K & & 1,071 & 200.0K \\
        & Technology & 916 & 1.3M & & 596 & 638.5K \\
        & Science & 538 & 343.6K & & 617 & 1.7M \\
        & Pooled & 2,931 & 2.4M & & 3,869 & 2.8M \\
        \bottomrule
    \end{tabular}
    }
    \caption[List of datasets used evaluation.]{Datasets used for evaluating XTR$_\text{base}$/WARP performance. The evaluation includes 6 datasets from BEIR \cite{beir} and 6 from LoTTE \cite{colbert2}.}
    \label{table:datasets_info}
\end{table}

\label{experimental_settings}

We now evaluate \WARP{} on six datasets from BEIR \cite{beir} and six datasets from LoTTE \cite{colbert2} listed in \Cref{table:datasets_info}. We use servers with 28 \texttt{Intel Xeon Gold 6132 @ 2.6 GHz} CPU cores\footnote{Each core has 2 threads for a total of 56 threads.} and 500 GB RAM. The servers have two NUMA sockets with roughly 92 ns intra-socket memory latency, 142 ns inter-socket memory latency, 72 GBps intra-socket memory bandwidth, and 33 GBps inter-socket memory bandwidth.

When measuring latency for end-to-end results, we compute the average latency of all queries and report the minimum average latency across three trials. For other results, we describe the specific measurement procedure in the relevant section. We measure latency on an otherwise idle machine. XTR's token retrieval stage does not benefit from GPU acceleration due to its use of ScaNN \cite{scann}, specifically designed for single-threaded\footnote{As of the recently released version 1.3.0, ScaNN supports multi-threaded search via the \texttt{search\_batched\_parallel} function.} use on x86 processors with AVX2 support. Therefore, unless otherwise specified, all measurements are performed on the CPU using a single thread\footnote{We believe \WARP{}'s design to naturally carry over to a GPU implementation, though this is left for future work.}.

\subsection{End-to-End Results}

Recent studies, such as those in \cite{plaidclusters}, demonstrate that dense multi-vector retrieval systems deliver near-exhaustive search performance while significantly reducing computational costs. In contrast, conventional re-ranking, though generally faster, often underperforms multi-vector retrieval systems in terms of effectiveness. As a result, we limit our evaluation to a comparison with the XTR/ScaNN baseline and leave a more extensive evaluation of re-ranking or exhaustive approaches to future work.

\Cref{table:lotte_success5} presents our results on LoTTE. XTR$_\text{base}$/WARP outperforms the optimized XTR$_\text{base}$/ScaNN implementation in terms of Success@5, while significantly reducing end-to-end latency, with speedups ranging from $4.6$x on LoTTE Lifestyle to $12.8$x on LoTTE Pooled. As shown in \Cref{table:beir_ncdg10}, we observe a similar trend with the evaluation of nDCG@10 on the six BEIR \cite{beir} datasets.
XTR$_\text{base}$/WARP achieves speedups of $2.7$x-$6$x over XTR$_\text{base}$/ScaNN with a slight gain in nDCG@10. Likewise, we find improvements of Recall@100 on BEIR with substantial gains in end-to-end latency, but we omit them here due to space constraints.

In comparison, EMVB \cite{emvb} reports a notable $2.9$x speedup in retrieval latency (142ms vs. 411ms) over ColBERTv2/PLAID on the LoTTE Pooled development set. By contrast, our method achieves a $4.3$x speedup (95ms vs. 405ms), albeit under different hardware settings and encoder models. Importantly, we view \WARP{} and EMVB as largely orthogonal approaches, suggesting that future work could explore integrating SIMD-based acceleration into WARP.

\begin{table}[t]
\centering
\resizebox{\columnwidth}{!}{
\begin{tabular}{@{}l|cccccc!{\vrule width 1pt}c@{}}
\toprule
& Lifestyle & Recreation & Writing & Technology & Science & Pooled & Avg. \\
\midrule
        
BM25 & 63.8 & 56.5 & 60.3 & 41.8 & 32.7 & 48.3 & 50.6\\
ColBERT & 80.2 & 68.5 & 74.7 & 61.9 & 53.6 & 67.3 & 67.7\\
GTR$_\text{base}$ & 82.0 & 65.7 & 74.1 & 58.1 & 49.8 & 65.0 & 65.8\\ \midrule
\rowcolor[gray]{0.90}XTR/ScaNN\footnotemark & \textbf{83.5} {\footnotesize(333.6)} & \textbf{69.6} {\footnotesize(400.2)} & 78.0 {\footnotesize(378.0)} & 63.9 {\footnotesize(742.5)} & 55.3 {\footnotesize(1827.6)} & 68.4 {\footnotesize(2156.3)} & 69.8\\
\rowcolor[gray]{0.90}XTR/WARP & \textbf{83.5} {\footnotesize(73.1)} & 69.5 {\footnotesize(72.4)} & \textbf{78.6} {\footnotesize(73.6)} & \textbf{64.6} {\footnotesize(96.4)} & \textbf{56.1} {\footnotesize(156.4)} & \textbf{69.3} {\footnotesize(171.3)} & \textbf{70.3} \\ \midrule
{\color{gray} Splade$_\text{v2}$ $^{\clubsuit\diamondsuit}$} & {\color{gray} 82.3} & {\color{gray} 69.0} & {\color{gray} 77.1} & {\color{gray} 62.4} & {\color{gray} 55.4} & {\color{gray} 68.9} & {\color{gray} 69.2}\\
{\color{gray} ColBERT$_\text{v2}$ $^{\clubsuit\diamondsuit}$} & {\color{gray} 84.7} & {\color{gray} 72.3} & {\color{gray} 80.1} & {\color{gray} 66.1} & {\color{gray} 56.7} & {\color{gray} 71.6} & {\color{gray} 71.9}\\ \midrule
{\color{gray} GTR$_\text{xxl}$} & {\color{gray} 87.4} & {\color{gray} 78.0} & {\color{gray} \underline{83.9} } & {\color{gray} 69.5} & {\color{gray} 60.0} & {\color{gray} 76.0} & {\color{gray} 75.8}\\
{\color{gray} XTR$_\text{xxl}$} & {\color{gray} \underline{89.1} } & {\color{gray} \underline{79.3} } & {\color{gray} 83.3} & {\color{gray} \underline{73.7} } & {\color{gray} \underline{60.8} } & {\color{gray} \underline{77.3} } & {\color{gray} \underline{77.3} }\\
\bottomrule
    \end{tabular}
    }
    \begin{tablenotes}
    \scriptsize
        \item $\clubsuit$: cross-encoder distillation \quad $\diamondsuit$: model-based hard negatives
    \end{tablenotes}
           
    \caption[Success@5 on LoTTE]{Success@5 on LoTTE. Numbers in parentheses show average latency (milliseconds), with the final column displaying the average score across the datasets. Both XTR/ScaNN and WARP use the XTR$_\text{base}$ model. XTR/ScaNN uses $k' = 40\,000$ and WARP uses $n_\text{nprobe} = 32$.}
    \label{table:lotte_success5}
\end{table}

\newcounter{mynote}
\setcounter{mynote}{\value{footnote}}

\begin{table}[t]
\centering
\resizebox{\columnwidth}{!}{
\begin{tabular}{@{}l|cccccc!{\vrule width 1pt}c@{}}
\toprule
& NFCorpus & SciFact & SCIDOCS & FiQA-2018 & Touché-2020 & Quora & Avg. \\
\midrule
        
BM25 & 32.5 & 66.5 & 15.8 & 23.6 & \underline{36.7}  & 78.9 & 42.3\\
ColBERT & 30.5 & 67.1 & 14.5 & 31.7 & 20.2 & 85.4 & 41.6\\
GTR$_\text{base}$ & 30.8 & 60.0 & 14.9 & 34.9 & 21.5 & 88.1 & 41.7\\ \midrule
\rowcolor[gray]{0.90}XTR/ScaNN\footnotemark[\value{mynote}] & \textbf{33.5} {\footnotesize(158.1)} & 69.6 {\footnotesize(309.7)} & 14.3 {\footnotesize(297.3)} & 34.1 {\footnotesize(338.2)} & \textbf{31.2} {\footnotesize(560.2)} & 86.0 {\footnotesize(411.2)} & 44.8\\
\rowcolor[gray]{0.90}XTR/WARP & \textbf{33.5} {\footnotesize(58.0)} & \textbf{70.5} {\footnotesize(64.3)} & \textbf{15.2} {\footnotesize(66.1)} & \textbf{34.2} {\footnotesize(70.7)} & 30.5 {\footnotesize(94.8)} & \textbf{86.2} {\footnotesize(67.6)} & \textbf{45.0} \\ \midrule
{\color{gray} Splade$_\text{v2}$ $^{\clubsuit\diamondsuit}$} & {\color{gray} 33.4} & {\color{gray} 69.3} & {\color{gray} 15.8} & {\color{gray} 33.6} & {\color{gray} 27.2} & {\color{gray} 83.8} & {\color{gray} 43.8}\\
{\color{gray} ColBERT$_\text{v2}$ $^{\clubsuit\diamondsuit}$} & {\color{gray} 33.8} & {\color{gray} 69.3} & {\color{gray} 15.4} & {\color{gray} 35.6} & {\color{gray} 26.3} & {\color{gray} 85.2} & {\color{gray} 44.3}\\ \midrule
{\color{gray} GTR$_\text{xxl}$} & {\color{gray} 34.2} & {\color{gray} 66.2} & {\color{gray} 16.1} & {\color{gray} \underline{46.7} } & {\color{gray} 23.3} & {\color{gray} \underline{89.2} } & {\color{gray} 45.9}\\
{\color{gray} XTR$_\text{xxl}$} & {\color{gray} \underline{35.3} } & {\color{gray} \underline{74.3} } & {\color{gray} \underline{17.1} } & {\color{gray} 43.8} & {\color{gray} 30.9} & {\color{gray} 88.1} & {\color{gray} \underline{48.3} }\\
\bottomrule
    \end{tabular}
    }
    \begin{tablenotes}
    \scriptsize
        \item $\clubsuit$: cross-encoder distillation \quad $\diamondsuit$: model-based hard negatives
    \end{tablenotes}
           
    \caption[nDCG@10 on BEIR]{nDCG@10 on BEIR. Numbers in parentheses show average latency (milliseconds), with the final column displaying the average score across the datasets. Both XTR/ScaNN and WARP use the XTR$_\text{base}$ model. XTR/ScaNN uses $k' = 40\,000$ and WARP uses $n_\text{nprobe} = 32$.}
    \label{table:beir_ncdg10}
\end{table}

\subsection{Scalability}

We now assess \WARP{}'s scalability in relation to both dataset size and the degree of parallelism.
To study the effect of the dataset size on \WARP{}'s performance, we evaluate its latency across development datasets of varying sizes: BEIR NFCorpus, BEIR Quora, LoTTE Science, LoTTE Technology, and LoTTE Pooled (\Cref{table:datasets_info}). \Cref{fig:scale_datasets} plots the  latency of different configurations versus the size of the dataset, measured in the number of document token embeddings. Our results confirm that \WARP{}'s latency generally scales  with the square root of the dataset size --- this is intuitive, as the number of clusters is by design proportional to the square root of the dataset size. 

A key advantage of \WARP{} over the reference implementation is its ability to leverage multi-threading. \Cref{fig:scale_threads} illustrates \WARP{}'s performance on the LoTTE Pooled development set, showing how the number of CPU threads impacts performance for different values of $n_\text{probe}$. Our results indicate that \WARP{} effectively parallelizes across multiple threads, achieving a speedup of $3.1$x for $n_\text{probe} = 32$ with 16 threads. We refer to \ref{appendix:latency_breakdowns}, and in particular \Cref{fig:warp_test_sets_mt}, for a more detailed breakdown of \WARP{}'s multi-threaded latency.

\footnotetext{Results differ marginally from \cite{xtr} due to float32 vs. bfloat16 encoder inference.}

\begin{figure}[t]
    \begin{subfigure}[t]{0.235\textwidth}
        \centering
        \includegraphics[width=\linewidth]{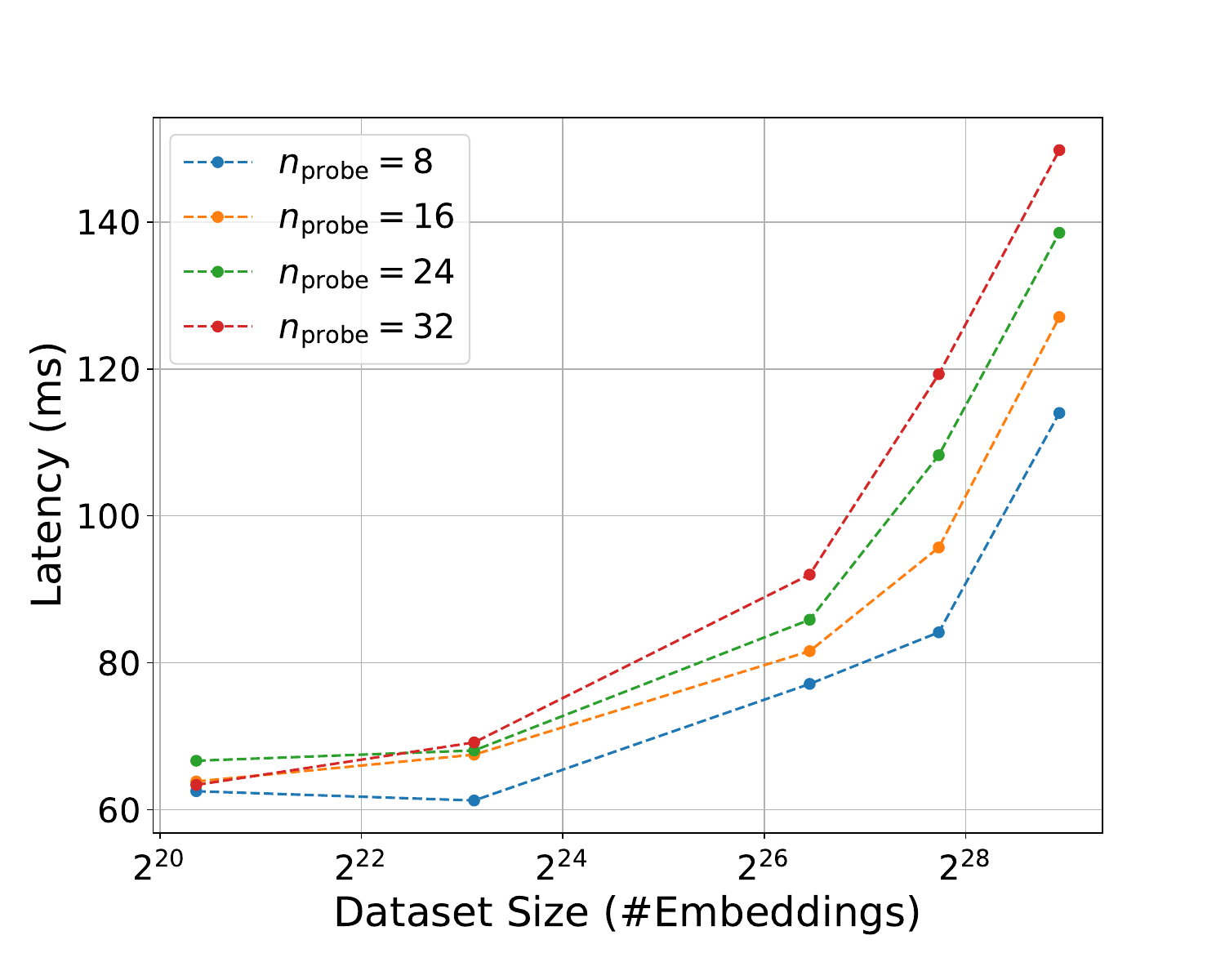}
        \caption[End-to-end latency as a function of dataset size]{End-to-end latency vs dataset size (measured in \#embeddings)}
        \label{fig:scale_datasets}
    \end{subfigure}
    \hfill
    \begin{subfigure}[t]{0.235\textwidth}
        \centering
        \includegraphics[width=\linewidth]{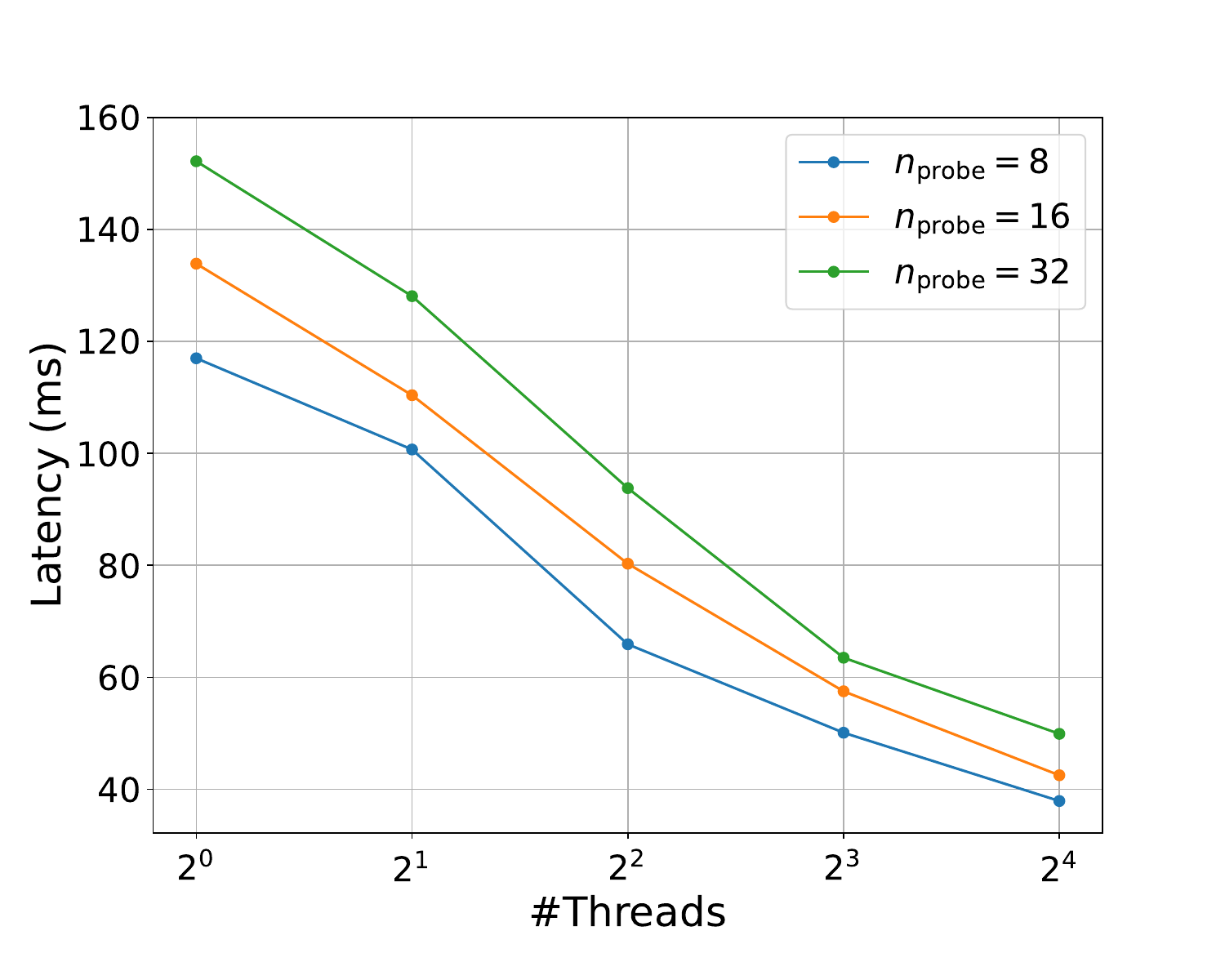}
        \caption{End-to-end latency for varying $n_\text{probe}$}
        \label{fig:scale_threads}
    \end{subfigure}
    \caption[\WARP{}'s scaling behavior with respect to the number of threads]{\WARP{}'s scaling behavior with respect to dataset size and the number of available CPU threads}
\end{figure}

\subsection{Memory Footprint}

\begin{table}[H]
    \centering
    \resizebox{\columnwidth}{!}{%
    \begin{tabular}{@{}llrrrrrr@{}}
        \toprule
        \multicolumn{3}{c}{} & \multicolumn{5}{c}{XTR} \\ 
        \multicolumn{2}{c}{\multirow{2}{*}{Dataset}} &  \multirow{2}{*}{\# Tokens} & \multicolumn{5}{c}{Index Size (GiB)} \\ \cmidrule{4-8}
        \multicolumn{2}{c}{} & & BruteForce & FAISS & ScaNN & WARP$_{(b = 2)}$ & WARP$_{(b = 4)}$ \\ \midrule

        \multirow{6}{*}{\textbf{BeIR}~\cite{beir} }
& NFCORPUS & 1.35M & 0.65 & 0.06 & 0.18 & 0.06 & 0.10 \\
& SciFact & 1.87M & 0.91 & 0.08 & 0.25 & 0.07 & 0.13 \\
& SCIDOCS & 6.27M & 3.04 & 0.28 & 0.82 & 0.24 & 0.43 \\
& Quora & 9.12M & 4.43 & 0.43 & 1.21 & 0.35 & 0.62 \\
& FiQA-2018 & 10.23M & 4.95 & 0.46 & 1.34 & 0.38 & 0.69 \\
& Touché-2020 & 92.64M & 45.01 & 4.28 & 12.22 & 3.40 & 6.16 \\

        \midrule
        \multirow{6}{*}{\textbf{LoTTE}~\cite{colbert2} }
& Lifestyle & 23.71M & 11.51 & 1.08 & 3.12 & 0.88 & 1.59 \\
& Recreation & 30.04M & 14.59 & 1.38 & 3.96 & 1.11 & 2.01 \\
& Writing & 32.21M & 15.64 & 1.48 & 4.25 & 1.19 & 2.15 \\
& Technology & 131.92M & 64.12 & 6.13 & 17.44 & 4.83 & 8.77 \\
& Science & 442.15M & 214.93 & 20.57 & 58.46 & 16.07 & 29.28 \\
& Pooled & 660.04M & 320.88 & 30.74 & 87.30 & 23.88 & 43.59 \\

\midrule\rowcolor[gray]{0.90}\multirow{1}{*}{\textbf{Total}} & -- & 1.44B & 700.66 & 66.98 & 190.52 & \underline{\textbf{52.48}} & 95.51 \\ \bottomrule

        \bottomrule
        \end{tabular}
    }
    \caption[Comparison of index sizes for the test datasets.]{Comparison of index sizes for the datasets.  Note that PLAID's memory usage is effectively identical to \WARP{}'s, only slightly larger.}
    \label{table:index_sizes_test}
\end{table}

\WARP{}’s primary advantage lies in its significant reduction in latency, though its benefits extend beyond speed alone. By adopting a ColBERTv2/PLAID-like approach for compression, \WARP{}'s advantage over XTR also translates into to a reduction in index size. This decrease in memory requirements broadens deployment options, particularly in resource-constrained environments. \Cref{table:index_sizes_test} compares index sizes across all evaluated test datasets. WARP$_{(b = 4)}$ demonstrates a substantially smaller index size compared to the BruteForce and ScaNN variants, providing a $7.3$x and $2$x reduction in memory footprint, respectively. Although the indexes generated by the FAISS implementation are marginally smaller, this comes at the cost of substantially reduced quality and latency. Notably, WARP$_{(b = 2)}$ outperforms the FAISS implementation in terms of quality with an even smaller index size\footnote{Additionally, \WARP{} reduces memory requirements compared to PLAID as it no longer requires storing a mapping from document ID to centroids/embeddings.}.

\section{Conclusion}

We present \WARP{}, a highly optimized engine for multi-vector retrieval built upon ColBERTv2/PLAID and the XTR framework. \WARP{} addresses key inefficiencies in existing systems through three major innovations: (1) \WARP{}$_\text{SELECT}$ for dynamic missing similarity imputation; (2) implicit decompression during retrieval; and (3) a streamlined two-stage reduction using dedicated C++ kernels. Together, these optimizations culminate in substantial performance gains, including a $41$x speedup over XTR on LoTTE Pooled -- cutting latency from over 6 seconds to just 171ms in single-threaded execution -- and a $3$x reduction in latency compared to ColBERTv2/PLAID, all without compromising retrieval quality. \WARP{} also scales efficiently with increased thread count and offers a significantly reduced memory footprint, making it well-suited for deployment in resource-constrained environments.

Future work may incorporate techniques such as SIMD and GPU acceleration, more lightweight query encoding, and end-to-end training with \WARP{} to better align model optimization with its retrieval process.

\printbibliography

\appendix
\clearpage

\section{Additional Results}

\subsection{Latency Breakdowns}
\label{appendix:latency_breakdowns}

\begin{figure}[t]
    \centering
    \includegraphics[width=0.45\textwidth]{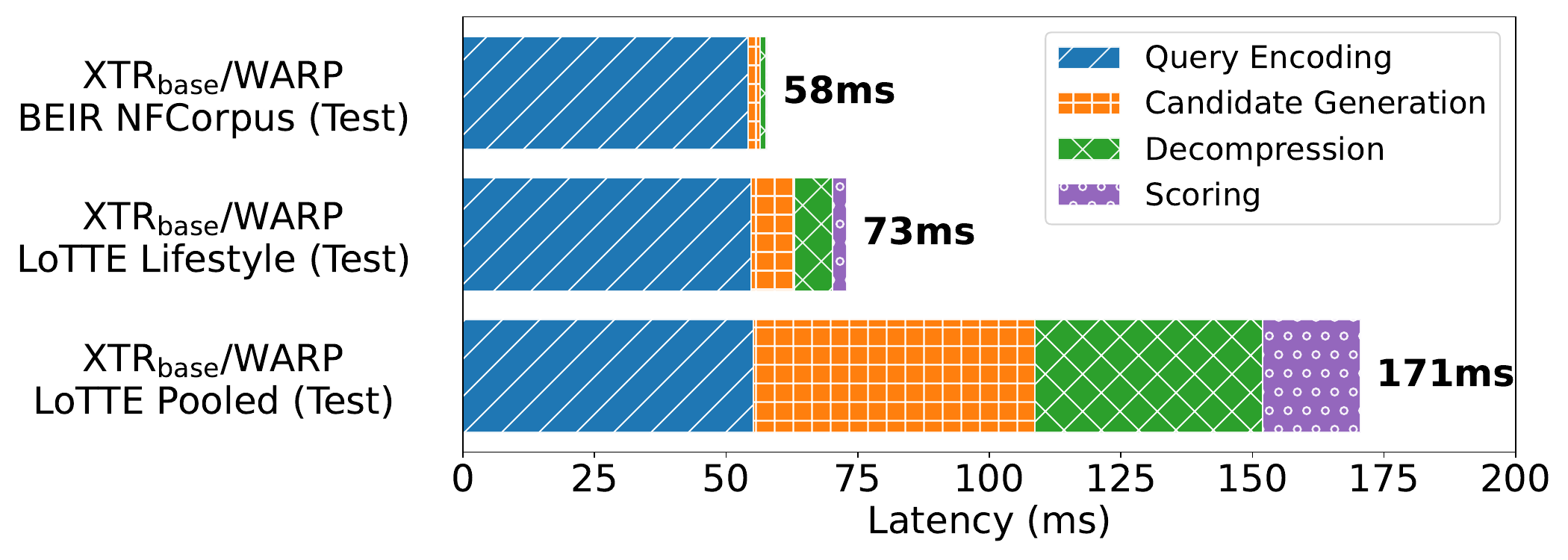}
    \caption[Breakdown of XTR$_\text{base}$/WARP's single-threaded latency]{Breakdown of XTR$_\text{base}$/WARP's avg. single-threaded latency for $n_\text{probe} = 32$ on the BEIR NFCorpus, LoTTE Lifestyle, and LoTTE Pooled datasets.}
\label{fig:warp_test_sets}
\end{figure}

In the following, we provide a more detailed breakdown of \WARP{}'s performance on three datasets of varying sizes: BEIR NFCorpus \cite{beir}, LoTTE Lifestyle \cite{colbert2}, and LoTTE Pooled \cite{colbert2}. \Cref{fig:warp_test_sets} illustrates the latency breakdown across four key stages: query encoding, candidate generation, decompression, and scoring.
For the smallest dataset, BEIR NFCorpus, the total latency is $58$ms, with query encoding dominating the process. Moving to the larger LoTTE Lifestyle dataset, the total latency increases to 73ms. Notably, on this dataset with over $100$K passages, \WARP{}'s entire retrieval pipeline -- comprising candidate generation, decompression, and scoring -- constitutes only about $25$\% of the end-to-end latency, with the remaining time spent on query encoding.
Even for the largest dataset, LoTTE Pooled, where the total latency reaches $171$ms, we observe that query encoding still consumes the majority of the processing time. While the other stages become more pronounced, query encoding remains the single most time-consuming stage of the retrieval process.
Without the use of specialized inference runtimes, query encoding accounts for approximately half of the execution time, thus presenting the primary bottleneck for end-to-end retrieval using \WARP{}.

\WARP{} is able to effectively parallelize execution over multiple threads. \Cref{fig:warp_test_sets_mt} shows the end-to-end latency breakdown for \WARP{} using $16$ threads. The decompression and scoring stages are fused in multi-threaded contexts. \WARP{} demonstrates great scalability, achieving substantial latency reduction across all stages. In the $16$-thread configuration, it notably surpasses the GPU-based implementation of PLAID on the LoTTE Pooled dataset.

\begin{figure}[t]
    \centering
    \includegraphics[width=0.45\textwidth]{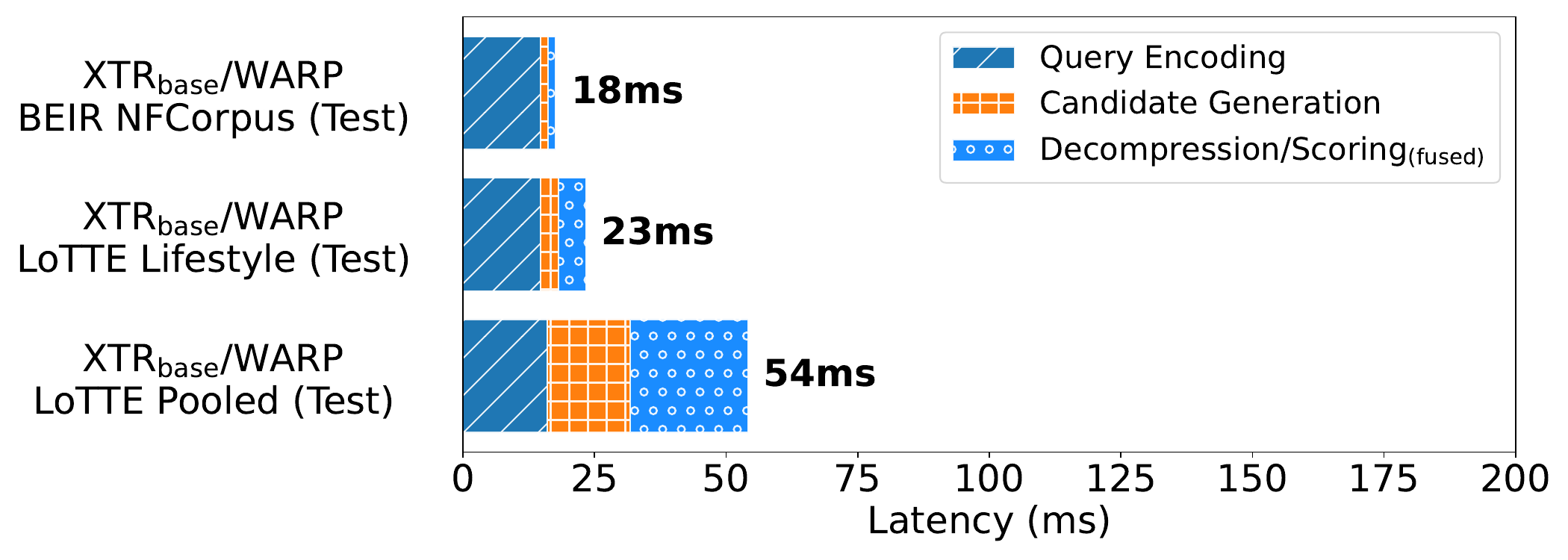}
    \caption[Breakdown of XTR$_\text{base}$/WARP's multi-threaded latency]{Breakdown of XTR$_\text{base}$/WARP's avg. latency for $n_\text{probe} = 32$ and $n_\text{threads} = 16$ on the BEIR NFCorpus, LoTTE Lifestyle, and LoTTE Pooled datasets}
    \label{fig:warp_test_sets_mt}
\end{figure}

\subsection{Performance Comparisons}

Similar to \Cref{fig:comp_lotte_pooled_test}, we analyze the performance of \WARP{} and contrast it with the performance of the baselines on the BEIR NFCorpus (\Cref{fig:comp_beir_nfcorpus_test}) and LoTTE Lifestyle (\Cref{fig:comp_lotte_lifestyle_test}) datasets. We find that \WARP{}'s single-threaded end-to-end latency is dominated by query encoding on BEIR NFCorpus and LoTTE Lifestyle, whereas the baselines introduce significant overhead via their retrieval pipelines.

\begin{figure}[t]
    \centering
    \includegraphics[width=0.45\textwidth]{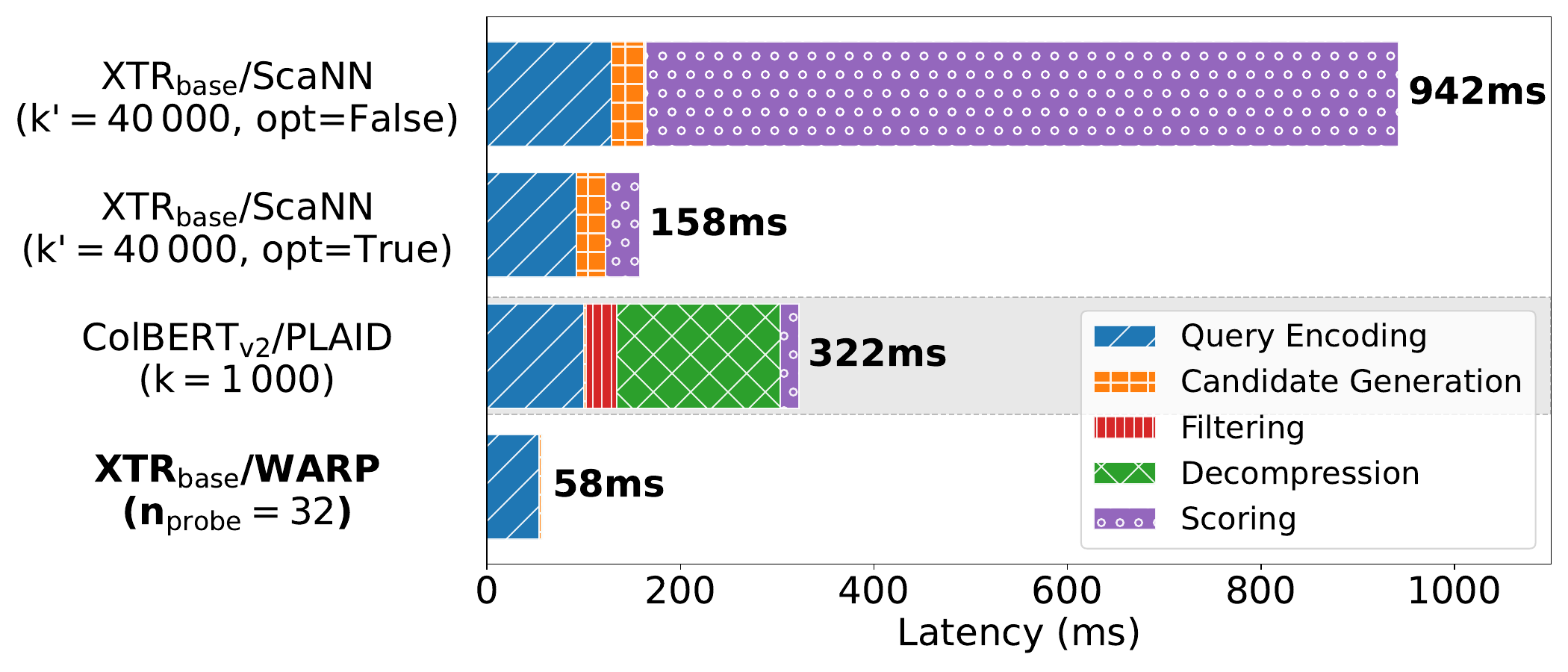}
    \caption[Latency breakdown comparison on BEIR NFCorpus]{Latency breakdown of the unoptimized reference implementation, optimized variant, ColBERTv2/PLAID, and XTR$_\text{base}$/WARP on BEIR NFCorpus Test}
    \label{fig:comp_beir_nfcorpus_test}
\end{figure}

\begin{figure}[t]
    \centering
    \includegraphics[width=0.45\textwidth]{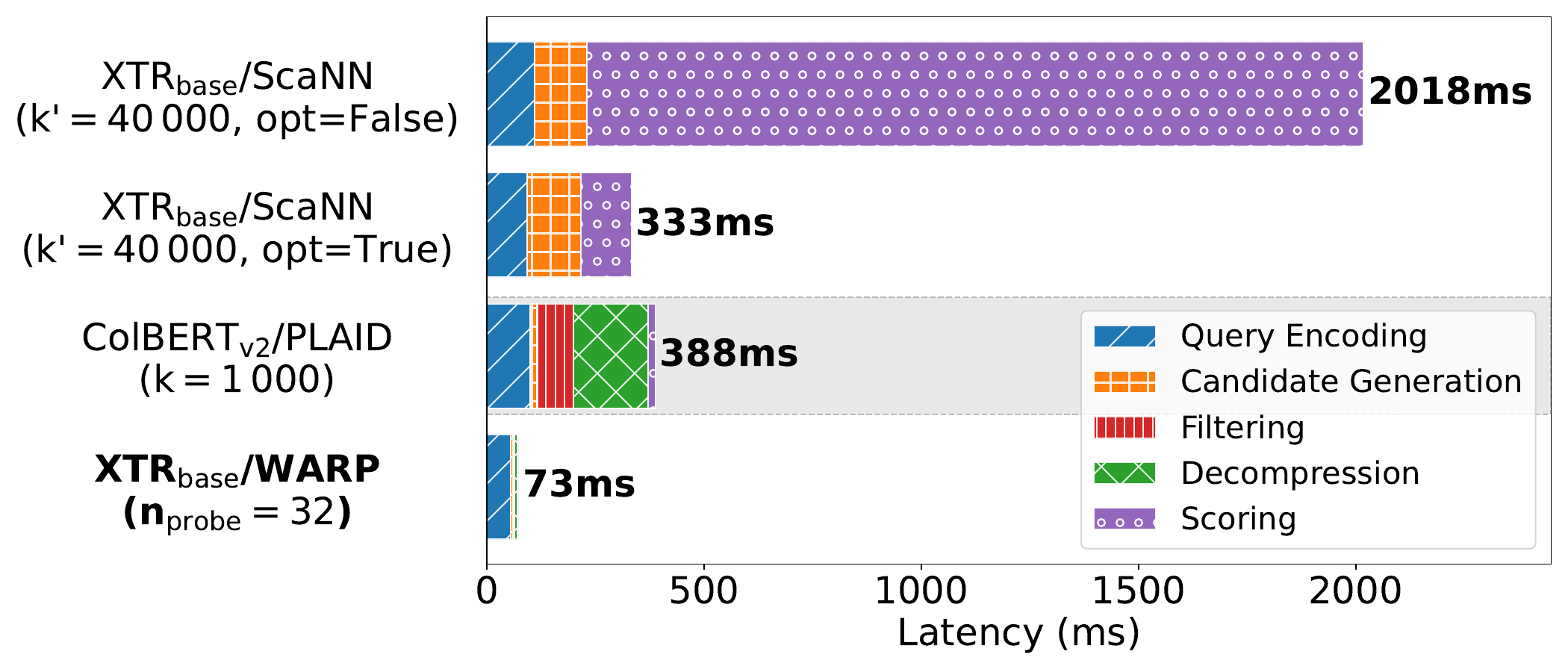}
    \caption[Latency breakdown comparison on LoTTE Lifestyle]{Latency breakdown of the unoptimized reference implementation, optimized variant, ColBERTv2/PLAID, and XTR$_\text{base}$/WARP on LoTTE Lifestyle Test}
    \label{fig:comp_lotte_lifestyle_test}
\end{figure}

\subsection{Evaluation of ColBERTv2/WARP}

To assess \WARP{}'s ability to generalize beyond the XTR model, we conduct experiments using ColBERTv2 in place of XTR$_\text{base}$ for query encoding. The results, presented in \Cref{colbertv2_vs_warp}, show that \WARP{} performs competitively with PLAID, despite not being specifically designed for retrieval with ColBERTv2. This suggests that \WARP{}'s approach may generalize effectively to retrieval models other than XTR. A detailed analysis of this generalization is deferred to future work.

\begin{table}[H]
\centering
\resizebox{\columnwidth}{!}{
\begin{tabular}{@{}l|cccccc!{\vrule width 1pt}c@{}}
\toprule
& NFCorpus & SciFact & SCIDOCS & FiQA-2018 & Touché-2020 & Quora & Avg. \\
\midrule
        
{\color{gray} ColBERT$_\text{v2}$/PLAID (k$=10$)} & {\color{gray} 33.3} & {\color{gray} 69.0} & {\color{gray} 15.3} & {\color{gray} 34.5} & {\color{gray} 25.6} & {\color{gray} 85.1} & {\color{gray} 43.8}\\
{\color{gray} ColBERT$_\text{v2}$/PLAID (k$=100$)} & {\color{gray} 33.4} & {\color{gray} 69.2} & {\color{gray} 15.3} & {\color{gray} 35.4} & {\color{gray} 25.2} & {\color{gray} 85.4} & {\color{gray} 44.0}\\
{\color{gray} ColBERT$_\text{v2}$/PLAID (k$=1000$)} & {\color{gray} 33.5} & {\color{gray} 69.2} & {\color{gray} 15.3} & {\color{gray} \underline{35.5} } & {\color{gray} 25.6} & {\color{gray} \underline{85.5} } & {\color{gray} 44.1}\\ \midrule
\rowcolor[gray]{0.90}ColBERT$_\text{v2}$/WARP (n$_\text{probe}=32$) & \underline{34.6}  & \underline{70.6}  & \underline{16.2}  & 33.6 & \underline{26.4}  & 84.5 & \underline{44.3} \\
\bottomrule
    \end{tabular}
    }   
    \caption[ColBERTv2/WARP nDCG@10 on BEIR]{ColBERTv2/WARP nDCG@10 on BEIR. The last column shows the average over 6 BEIR datasets.}
    \label{colbertv2_vs_warp}
\end{table}

\end{document}